\documentclass[authoryear,3p,11pt]{elsarticle}

\usepackage{longtable}
\usepackage{amsmath,amssymb}
\usepackage{graphicx}

\usepackage{amssymb}
\usepackage{mathtools}

\usepackage{lipsum}
\usepackage{breqn}

\begin{document}

\begin{frontmatter}

\title{Onset of oligarchic growth and implication 
for accretion histories of dwarf planets}

\author[ucla,jpl]{Ryuji Morishima\corref{cor1}}
\ead{Ryuji.Morishima@jpl.nasa.gov}

\cortext[cor1]{Corresponding author}

\address[ucla]{University of California, Los Angels, Institute of Geophysics and Planetary Physics, Los Angeles, CA  90095, USA}
\address[jpl]{Jet Propulsion Laboratory/California Institute of Technology,
Pasadena, CA 91109, USA}

\begin{abstract}
We investigate planetary accretion 
that starts from equal-mass planetesimals 
using an analytic theory and numerical simulations.
We particularly focus on how the planetary mass $M_{\rm oli}$ at the onset of oligarchic growth
depends on the initial mass $m_0$ of a planetesimal.
Oligarchic growth commences when the velocity dispersion relative to the Hill velocity of the protoplanet takes its minimum.
We find that if $m_0$ is small enough, 
this normalized velocity dispersion becomes as low as unity during 
the intermediate stage between the runaway and oligarchic growth stages. 
In this case, $M_{\rm oli}$ is independent of $m_0$.
If  $m_0$ is large, on the other hand, oligarchic growth commences directly after 
runaway growth, and $M_{\rm oli} \propto m_0^{3/7}$.
The planetary mass $M_{\rm oli}$ for the solid surface density of the Minimum Mass Solar Nebula is close to the masses of the dwarf planets
in a reasonable range of $m_0$. This indicates that  they are likely to be the largest remnant planetesimals that failed to become planets.
The power-law exponent $q$ of the differential mass distribution of remnant planetesimals 
is typically $-2.0$ and $-2.7$ to $-2.5$ for small and large $m_0$.
The slope, $q \simeq -2.7$,  
and the bump at $10^{21}$ g (or 50 km in radius) for the mass distribution of hot Kuiper belt objects 
are reproduced if $m_0$ is the bump mass. 
On the other hand, small initial planetesimals with $m_0 \sim 10^{13}$ g or less
are favored to explain the slope of large asteroids, $q \simeq -2.0$, while the bump at 
$10^{21}$ g can be reproduced by introducing a small number of 
asteroid seeds each with mass of $10^{19} $ g.
\end{abstract}


\end{frontmatter}

\section{Introduction}
The standard scenario of planet formation begins with small dust floating in a gaseous disk.  
Clumping of dust particles either due to streaming instability \citep{Johansen2007, Johansen2015a},  
turbulent concentration \citep{Cuzzi2008,Cuzzi2010,Chambers2010,Hopkins2016}, 
or direct sticking \citep{Okuzumi2012,Windmark2012, Kataoka2013,Garaud2013}
leads to formation of gravitationally bound planetesimals. 
These planetesimal formation models generally predict the size of initial planetesimals 
ranging from 10 to 100 km or even larger although  some direct sticking models \citep{Windmark2012,Garaud2013} 
predict planetesimals as small as 100 m.
The initial planetesimal size implied from planetary accretion models is controversial.
For models of asteroid formation, 
\citet{Morbidelli2009} suggested  large size ($\sim$ 100 km) of initial planetesimals 
whereas \citet{Weidenschilling2011} showed that the asteroid size distribution 
can also be reproduced from initially small ($\sim$ 100 m) planetesimals.
Accretion models for Kuiper belt objects (KBOs)  generally 
favor small initial planetesimals (1 -10 km) since the accretion timescale is very long 
for initially large planetesimals \citep{Kenyon2012,Schlichting2013}.

Once planetesimals form, they further grow via mutual collisions.
In the early stage of planetary accretion, called the runway growth stage, large planetesimals grow quickly 
due to their gravitational focusing effects while small planetesimals nearly remain at their initial sizes 
\citep{Greenberg1978,Wetherill1989,Kokubo1996,Barnes2009}. 
As growth proceeds,  the largest bodies, often called planetary embryos or protoplanets,
start to dominate the gravitational scattering effect, i.e. viscous stirring \citep{Ida1993}.
As the number of protoplanets decreases, their orbital separation increases.
Eventually, each small body is predominantly stirred by a single protoplanet \citep{Kokubo1998}.
Because of the viscous stirring effect of each protoplanet, among protoplanets, larger ones grow more slowly than smaller ones.
As a result, protoplanets with similar masses grow at a similar rate \citep{Kokubo1998, Kokubo2002}.
This growth mode is called oligarchic growth.
Although the basic pictures of runaway and oligarchic growth are apparently well known, 
the pathway from runaway growth to oligarchic growth described in the literature varies from author to author.
The mass of a protoplanet $M_{\rm oli}$ at the onset of oligarchic growth is not well understood either.

Based on $N$-body simulations and the analytic formulation, 
\citet{Ida1993} derived the condition that 
protoplanets rather than small planetesimals dominate viscous stirring.
They argued that once this condition is fulfilled, transition from runaway growth to oligarchic growth occurs.
As pointed out by \citet{Ormel2010a}, however, this condition is already satisfied even during runaway growth and is 
thus necessary but not sufficient for oligarchic growth.

\citet{Kokubo1998} performed direct $N$-body simulations of oligarchic growth and 
found that the orbital separation of neighboring protoplanets is about 10 Hill radii as a result of orbital repulsion.
Since this width is the typical scale of gravitational influence of a protoplanet, this single protoplanet  
dominates viscous stirring in the region of its gravitational influence. 
In earlier stages of planetary accretion, the mutual separation between neighboring protoplanets  
is smaller than 10 Hill radii and their regions of gravitational influence mutually overlap.  

\citet{Ormel2010a} derived a new condition for the transition from runaway to oligarchic growth. 
They examined evolution of the velocity dispersion $u$ of small planetesimals normalized by 
the Hill velocity $v_{\rm H}$ of the largest protoplanet, 
where $v_{\rm H}$ is the product of Hill radius of the protoplanet and the Keplerian frequency. 
The normalized velocity $u/v_{\rm H}$ decreases during the runaway growth while it increases during oligarchic growth.  
The transition between these two stages then occurs when $u/v_{\rm H}$ takes its minimum. 
Based on timescale arguments for the dispersion dominant regime ($u/v_{\rm H}> 1$), 
they derived the planetary mass $M_{\rm oli}$  at the transition. 
They showed that $M_{\rm oli} \propto m_0^{3/7}$, where $m_0$ is the initial mass of each planetesimal.
\citet{Morishima2013} derived a similar transition mass 
without timescale arguments assuming that
the protoplanet is the largest body of a continuous 
mass distribution with the power-law exponent $q$ of $-2.5$, 
which is the typical value during runaway growth 
\citep{Kokubo1996,Ormel2010b,Morishima2013}.

Using the simple analytic formulation originally developed by \citet{Goldreich2004},  
\citet{Lithwick2014} showed that runaway growth is followed by a new stage - the trans-Hill stage - during which 
$u/v_{\rm H}$ remains nearly unity. 
The growth rates of protoplanets during the trans-Hill stage are
independent of mass and that leads to $q \simeq -2.0$ for the massive side of the mass distribution. 
The trans-Hill stage is terminated by emergence of oligarchic growth, 
in which the mutual separation between protoplanets is scaled by the Hill radius \citep{Kokubo1998}. 
The planetary mass $M_{\rm oli}$ at the onset of oligarchic growth subsequent to the trans-Hill stage 
is independent  of $m_0$.
The concept of the trans-Hill stage was confirmed by coagulation simulations
of \citet{Shannon2015}.

\citet{Schlichting2011} also discussed the transition stage after runaway growth using 
the analytic formulation of \citet{Goldreich2004}. 
They also found that $u/v_{\rm H}$ takes a fixed value close to but somewhat larger than unity 
during the transition stage, contrary to \citet{Lithwick2014}. 
The difference comes from their assumptions; 
while \citet{Schlichting2011} assumed that growth of protoplanets  is equally contributed
by merging with small planetesimals and mutual merging between protoplanets (equal accretion), 
\citet{Lithwick2014} assumed that growth of protoplanets is dominated by merging with small planetesimals.
\citet{Lithwick2014}, in fact, analytically proved his assumption and criticized the unjustified assumption
employed by \citet{Schlichting2011}. 

As reviewed, different authors showed apparently different conditions for or pathways to the onset of oligarchic growth. 
The objective of the present paper is to establish a comprehensive picture 
for the onset of oligarchic growth and to derive  $M_{\rm oli}$ for arbitrary $m_0$, using 
an analytic formulation and numerical simulations.  
Particularly, we will clarify the following three points.
\begin{enumerate}
\item Three conditions for the onset of oligarchic growth were indicated in the literature:
\begin{enumerate}
\item protoplanets dominate viscous stirring, 
\item the mutual separation between neighboring protoplanets is 10 Hill radii, and
\item $u/v_{\rm H}$ takes its minimum.
\end{enumerate}
The condition (b) is equivalent to the definition of oligarchic growth. 
We will show that the condition (c) is useful to pin down the timing of the onset of oligarchic growth, 
as shown by \citet{Ormel2010a}, although the condition (a) is also required for oligarchic growth. 

\item \citet{Ormel2010a} and \citet{Morishima2013} showed that both $M_{\rm oli}$ and the minimum $u/v_{\rm H}$ depend on $m_0$ while 
\citet{Schlichting2011} and \citet{Lithwick2014} showed that  $M_{\rm oli}$ and the minimum $u/v_{\rm H}$ are both independent of $m_0$.
We will show that the formulations of \citet{Ormel2010a} and \citet{Morishima2013} are applicable for large  $m_0$ 
while the formulations of \citet{Schlichting2011} and \citet{Lithwick2014} are applicable for small $m_0$.
The critical value of $m_0$  which separates these two regimes will be explicitly formulated. 
The transition stage between the runway growth and oligarchic growth stages appears only if 
$m_0$ is lower than this critical value.

\item \citet{Schlichting2011} and \citet{Lithwick2014}  employed different assumptions 
for the contribution of mutual merging between protoplanets to their growth. 
We will show that as long as their equations are employed, 
growth of protoplanets is dominated by merging with small planetesimals, as proved by \citet{Lithwick2014}.
We will, however, show that 
the contribution of mutual merging can be comparable to or even larger than 
the small bodies' contribution due to 
very small inclinations of large bodies and 
to small bodies' velocities somewhat larger than the Hill velocity of large bodies.

\end{enumerate}

Besides clarifying the picture of the onset of oligarchic growth,
at least, the following two important implications are discussed from $M_{\rm oli}$.
First, the masses of the dwarf planets are potentially close to $M_{\rm oli}$.
Protoplanets more massive than $M_{\rm oli}$ have low velocity dispersion due to dynamical friction of 
surrounding planetesimals. As a result, they efficiently merge together and form large planets, while 
leaving remnant planetesimals with the largest one about  $M_{\rm oli}$ in mass \citep{Morishima2008}. 
This argument is probably applied to asteroids.
For KBOs at large heliocentric distances, 
planetary accretion is likely to be incomplete.
Recent observations of KBOs \citep{Fraser2014,Adams2014} showed that 
the largest two bodies, Pluto and Eris, are distinctively separated from a continuous size distribution,
possibly indicating that they actually entered into oligarchic growth but their growth was stalled immediately after that.
In either case, it is worthwhile to compare the masses of the dwarf planets in the asteroid and Kuiper belts 
with $M_{\rm oli}$ as it helps us to infer the physical parameters of the proto-solar nebula.
In 2015, NASA's Dawn and New Horizons spacecraft arrived at Ceres \citep{Russell2015} and Pluto \citep{Stern2015}, respectively.
The present study may be able to give a hint for why they failed to become planets.
Additional constraints on their accretion histories are obtained from the mass distribution slopes of asteroids and KBOs.

Second, rapid migration of protoplanets caused by gravitational interactions with surrounding planetesimals 
is likely to occur once their masses reach $\sim M_{\rm oli}$.
A protoplanet gravitationally scatters surrounding planetesimals. 
Back reaction on the protoplanet causes its radial migration, called planetesimal-driven migration (PDM), 
as the exerted torques do not cancel out 
\citep{Ida2000,Kirsh2009,Bromley2011,Ormel2012,Kominami2016}.
PDM is important for planet formation as it can significantly modify
planetary growth rates and mutual separations between planets \citep{Levison2010}.
\citet{Minton2014} pointed out that the protoplanet must be much more massive than
other nearby bodies to trigger smooth PDM. Otherwise, stochastic forces from nearby bodies prohibit smooth PDM. 
They found that such a condition is generally fulfilled at large distance from the central star ($>$ 1 AU) and with the protoplanet's
mass of about the lunar mass, although the initial planetesimal size was  limited to $\sim$ 100 km in their study.
The condition of mass separation is expected to be fulfilled at best at the onset of oligarchic growth, 
since the largest protoplanet is caught up with by other protoplanets in mass in the midst of oligarchic growth. 
Particle-based hybrid codes \citep{Levison2012,Morishima2015} for planetary accretion developed 
recently can handle PDM of protoplanets.  The lowest protoplanet's mass capable in these codes
is the mass of a tracer,  which represents a large number of planetesimals. 
Thus, $M_{\rm oli}$ is a good indicator for 
the tracer mass that should be used in these new codes. 

In Section~2, we develop an analytic theory for planetary growth and 
derive the planetary mass $M_{\rm oli}$ at the onset of oligarchic growth for arbitrary initial mass of 
each planetesimal $m_0$.
In Section~3, we perform numerical simulations to check the validity of our theory 
using a pure $N$-body code \citep{Morishima2010} and a particle-based hybrid code \citep{Morishima2015}.
Implications of the present study and effects not considered in Sections~2 and 3 are discussed in Section~4.
The summary is given in Section~5.

\section{Theory}
We employ the two-component model developed by \citet{Goldreich2004}
and use their notation as much as possible. 
Consider a system of small and large bodies with the surface densities of $\sigma$ and $\Sigma$, respectively.
The mass, radius, and velocity dispersion of the 
 small bodies are $m_0$, $s$, and $u$ and 
 those for large bodies are $M$, $R$, and $v$.
All the bodies have the bulk density $\rho$. 
This two component approximation is easily extended to a continuous size distribution case.

The Hill radius of a large body  is 
\begin{equation}
R_{\rm H} = a h_{M} = a\left(\frac{M}{3M_{\odot}}\right)^{1/3},
\end{equation}
where 
$a$ is the semimajor axis of the body, $h_{M}$ is the reduced Hill radius,
and $M_{\odot}$ is the mass of the central star which we assume to be the solar mass.
The ratio of the physical radius to the Hill radius is defined as 
\begin{equation}
\alpha = \frac{R}{R_{\rm H}}  = 6.0 \times 10^{-3}\left(\frac{a}{1 \hspace{0.3em}{\rm AU}}\right)^{-1}\left(\frac{\rho}{2 \hspace{0.3em}{\rm g}\hspace{0.3em}{\rm cm}^{-3} }\right)^{-1/3}.
\end{equation}
The Hill velocity of the large body is given as
\begin{equation}
v_{\rm H} = R_{\rm H} \Omega, 
\end{equation}
where $\Omega$ is the Keplerian frequency at $a$. 

The  change rate of the velocity dispersion $u$ due to large bodies' viscous stirring  is 
\begin{equation}
\frac{1}{u}\frac{\mathrm{d}u}{\mathrm{d}t}  = \frac{\Sigma \Omega}{\rho R} \alpha^{-2} 
\left\{ \begin{array}{ll} 
(v_{\rm H}/u)^4
& \mbox{(if $u > v_{\rm H}  $)}, \\ 
v_{\rm H}/u
& \mbox{(if $ u < v_{\rm H}$)}.
\end{array}\right.
\label{eq:dudt}
\end{equation}
Here we ignored collisional damping and  gas drag. These effects are generally unimportant unless 
$s$ is very small (see Section~4.5).

For a more general continuous size distribution, the  change rate of $u$ due to all the bodies'  viscous stirring
is obtained if we replace $\Sigma$ by $M_{\rm stir}\sigma/M$ in Eq.~(\ref{eq:dudt}). Here
$M_{\rm stir}$ is the effective mass for viscous stirring \citep{Ida1993,Ormel2010a} given as 
\begin{equation}
M_{\rm stir} = \frac{\langle m^2 \rangle}{ \langle m \rangle} = -\frac{q+2}{q+3}\left(\frac{M}{m_0}\right)^{q+2}M, 
\label{eq:mstir}
\end{equation}
where the brackets
represent averaging over mass $m$ between $m_0$ and $M$. 
The latter equivalence in Eq.~(\ref{eq:mstir}) is applied 
if the mass distribution is given by a single power-law distribution $dN_c = m^q dm$, 
where $N_c$ is the cumulative number of the bodies in the system and $q$ is the power-law exponent. 
We also assumed that $-3< q < -2$. This range of $q$ appears most commonly in planetary accretion around the onset of oligarchic growth,  
although $M_{\rm stir}$ for other ranges of $q$ can be derived easily.
If $q > -3$, large bodies dominate viscous stirring \citep{Ida1993}.
This is expressed as 
\begin{equation}
M\Sigma \simeq M_{\rm stir}\sigma,
\label{eq:oli}
\end{equation} 
and we recover Eq.~(\ref{eq:dudt}).

\subsection{Growth due to merging with small bodies}
We first consider growth of large bodies due to merging with small bodies.
The contribution of mutual merging between large bodies is discussed in the next section.
The growth rate of the large body is 
\begin{equation}
\frac{1}{R}\frac{\mathrm{d}R}{\mathrm{d}t}  = \frac{\sigma \Omega}{\rho R} \alpha^{-1} 
\left\{ \begin{array}{ll} 
(v_{\rm H}/u)^2
& \mbox{(if $ v_{\rm H} < u <  \alpha^{-1/2}v_{\rm H} $)}, \\ 
v_{\rm H}/u
& \mbox{(if $\alpha^{1/2}v_{\rm H} < u < v_{\rm H}$)},
\end{array}\right.
 \label{eq:dr1dt}
\end{equation}
where $ \alpha^{-1/2}v_{\rm H}$ is about the escape velocity of the largest body.

The ratio $\Sigma/\sigma$ is called the efficiency \citep{Schlichting2011,Lithwick2014}. 
Provided that the normalized velocity dispersion, $u/u_{\rm H}$, evolves quasi-stationary, 
the relationship between the efficiency $\Sigma/\sigma$ and the velocity dispersion $u/v_{\rm H}$ 
at a certain time is derived 
by adopting $\mathrm{d} (u/v_{\rm H})/\mathrm{d}t = 0$ or equivalently by equating Eq.~(\ref{eq:dudt}) with Eq.~(\ref{eq:dr1dt}) as 
\begin{equation}
\frac{\Sigma}{\sigma} =  \alpha \left(\frac{u}{v_{\rm H}}\right)^2.
\label{eq:effi}
\end{equation}

As we will describe in detail below, 
planetary accretion is classified into three distinct stages.
\begin{enumerate}
\item  Runaway growth: $\mathrm{d}(\Sigma/\sigma)/\mathrm{d}M < 0$, $q < -2.0$, and $\mathrm{d}(\dot{R}/R)/ \mathrm{d}R > 0$.
\item  Trans-Hill  (neutral): $\mathrm{d}(\Sigma/\sigma)/\mathrm{d}M = 0$, $q =-2.0$, and $\mathrm{d}(\dot{R}/R)/ \mathrm{d}R = 0$.
\item  Oligarchic growth: $\mathrm{d}(\Sigma/\sigma)/\mathrm{d}M > 0$, $q > -2.0$, and $\mathrm{d}(\dot{R}/R)/ \mathrm{d}R < 0$.
\end{enumerate}
Here the index $q$ is used for largest bodies only. 

During the runaway growth, $\Sigma/\sigma (\simeq M_{\rm stir}/M)$ and $u/v_{\rm H}$ decrease with increasing $M$,
provided that $q$ changes only slowly (Eq.~(\ref{eq:mstir})). Typically, $q \simeq -2.7$ to $-2.5$ \citep{Kokubo1996,Ormel2010a}. 
 In other words, $v_{\rm H}$ (or $R$) grows faster than $u$ does. 
Once the ratio $u/v_{\rm H}$ becomes as low as unity, 
the change rates of $u$ and $v_{\rm H}$ (or $R$) become the same. 
As a result, $u/v_{\rm H}$ is kept to be nearly unity. 
The growth rates of large bodies are independent of mass (neutral growth) 
as long as their Hill velocities are larger than
the velocity dispersion. Thus, this stage, named the trans-Hill stage \citep{Lithwick2014}, 
is technically different from the traditional runaway growth stage,
in which larger bodies grow faster. 
During this stage, $q$ is expected to increase toward $-2$ as indicated by a constant $\Sigma/\sigma$ and neutral growth.
The trans-Hill stage is terminated by emergence of oligarchic growth. 

The above discussion for the trans-Hill stage is applied only if $m_0$ is small enough. 
For large $m_0$,  oligarchic growth commences directly after runaway growth before $u/v_{\rm H}$ reaches to unity 
in the absence of the trans-Hill stage, as shown below.

Oligarchic bodies gravitationally repulse each other and their mutual distance is kept to be about $b_e$ times the mutual Hill radii,
where $b_e \sim 10$ \citep{Kokubo1998}.
At the end of oligarchic growth, all the nearby planetesimals are accreted by protoplanets. 
The final mass of an oligarchic body is  
called the isolation mass $M_{\rm iso}$ \citep{Kokubo2002} given as 
\begin{equation}
M_{\rm iso} = 2\pi b_e a (2^{1/3}R_{\rm H}) \sigma = (2\pi b_e a^2  \sigma)^{3/2}\left(\frac{2}{3M_{\odot}} \right)^{1/2},  \label{eq:miso}
\end{equation} 
where $2^{1/3}R_{\rm H}$ is the mutual Hill radius and  $\sigma$ is the value at the beginning.
Using $M_{\rm iso}$ and Eq.~(\ref{eq:effi}),  the mass of the oligarchic body in the midst of oligarchic growth is given as  
\begin{equation}
M =  \left(\frac{\Sigma}{\sigma}\right)^{3/2}M_{\rm iso} = \alpha^{3/2} \left(\frac{u}{v_{\rm H}}\right)^3 (q+3)^{3/2}M_{\rm iso},
\label{eq:mem}
\end{equation} 
where the effect of depletion of small bodies is ignored.

If $m_0$ is small enough, the largest body's mass at the transition from  trans-Hill  to oligarchic growth is 
given as
\begin{equation}
M_{\rm oli,SS} =  \alpha^{3/2} M_{\rm iso}, \label{eq:molis}
\end{equation} 
as $u/v_{\rm H} = 1$ and $q \simeq -2$. The same transition mass except for the factor was derived by \citet{Lithwick2014}.
The efficiency at the transition for small $m_0$ is 
\begin{equation}
\left(\frac{\Sigma}{\sigma}\right)_{\rm SS}  = \alpha. \label{eq:sig0s}
 \end{equation}
 
If $m_0$ is not small enough, 
the largest body's mass at the transition from runaway  to oligarchic growth  is 
given as 
\begin{equation}
M_{\rm oli,L}(q) = \left[\left(-\frac{q+2}{q+3}\right)^{3} m_0^{-3(q+2)}M_{\rm iso}^{2}\right]^{-1/(3q+4)}, \label{eq:molilq}
\end{equation} 
where we used Eqs.~(\ref{eq:mstir}), (\ref{eq:oli}), and (\ref{eq:mem}). 
If $q = -2.5$, we have  
\begin{equation}
M_{\rm oli,L}(q = -2.5) =  m_0^{3/7}M_{\rm iso}^{4/7}. \label{eq:molil}
\end{equation} 
The same equation except for the factor was derived by \citet{Morishima2013}. 
The transition mass derived by \citet{Ormel2010a} is also similar to but slightly different 
from Eq.~(\ref{eq:molil}) since they employed different assumptions.
The efficiency and the lowest  $u/v_{\rm H}$ at the transition for large $m_0$ are respectively given as 
\begin{equation}
\left(\frac{\Sigma}{\sigma}\right)_{\rm L} =  \left(-\frac{q+3}{q+2}\right)^{2}  \left(\frac{m_0}{M_{\rm iso}}\right)^{2(q+2)/(3q+4)},
 \end{equation}
 and
 \begin{equation}
\left(\frac{u}{v_{\rm H}}\right)_{\rm L} =  \alpha^{-1/2} \left(\frac{\Sigma}{\sigma}\right)_{\rm L}^{1/2}
\label{eq:uminl}
\end{equation}
The critical planetesimal mass that separates these two regimes is given by equating Eq.~(\ref{eq:molis}) with Eq.~(\ref{eq:molil}) as 
$\alpha^{7/2} M_{\rm iso}$ for $q = -2.5$.

\subsection{Corrections due to merging between large bodies}

Now we consider the contribution of merging between large bodies to their growth.
Provided that $u \ge v_{\rm H}$,
the accretion rate of large bodies is given as
\begin{dmath}
\frac{1}{R}\frac{\mathrm{d}R}{\mathrm{d}t}  = \frac{\sigma \Omega}{\rho R} \alpha^{-1} \left(\frac{v_{\rm H}}{u}\right)^2
+  \frac{\Sigma \Omega}{\rho R} \alpha^{-1}  
\left\{ \begin{array}{ll} 
(v_{\rm H}/v)^2
& \mbox{(if $ v_{\rm H} < v < \alpha^{-1/2}v_{\rm H}$)}, \\ 
v_{\rm H}/v
& \mbox{(if $\alpha^{1/2}v_{\rm H} < v < v_{\rm H}$)},\\
\alpha^{-1/2}
& \mbox{(if $v < \alpha^{1/2}v_{\rm H} $)},
\end{array}\right.
 \label{eq:dr1dtc}
\end{dmath}
 where the first and the second terms in the r.h.s are the contributions from merging with small and large bodies, respectively.
The large bodies' velocity $v$ is determined by the balance between mutual stirring and dynamical friction due to small bodies: 
\begin{equation}
\frac{1}{v}\frac{\mathrm{d}v}{\mathrm{d}t}  = - \frac{\sigma \Omega}{\rho R} \alpha^{-2}  \left(\frac{v_{\rm H}}{u}\right)^4
+  \frac{\Sigma \Omega}{\rho R} \alpha^{-2} 
\left\{ \begin{array}{ll} 
(v_{\rm H}/v)^4
& \mbox{(if $ v_{\rm H} < v $)}, \\ 
v_{\rm H}/v
& \mbox{(if $ v < v_{\rm H}$)}.
\end{array}\right.
 \label{eq:dvdt}
\end{equation}
The velocity $v$ at the equilibrium state ($\mathrm{d}v/\mathrm{d}t = 0$) is 
\begin{equation}
\frac{v}{v_{\rm H}} = 
\left\{ \begin{array}{ll} 
\alpha^{1/4} (u/v_{\rm H})^{3/2} 
& \mbox{(if $ v_{\rm H} < v $)}, \\ 
\alpha (u/v_{\rm H})^{6} 
& \mbox{(if $ v < v_{\rm H}$)},
\end{array}\right.
 \label{eq:veq}
\end{equation}
where we used Eq.~(\ref{eq:effi}). 
The ratio of the large bodies' contribution to the small bodies' contribution 
is then given as 
\begin{equation}
\frac{\mathrm{d}R/\mathrm{d}t|_{\rm large}} {\mathrm{d}R/\mathrm{d}t|_{\rm small}}      = 
\left\{ \begin{array}{ll} 
1
& \mbox{(if $ v = u = \alpha^{-1/2}v_{\rm H}$)}, \\ 
\alpha^{1/3}  
& \mbox{(if $ v = v_{\rm H}$ and $u = \alpha^{-1/6} v_{\rm H}$)},\\
\alpha^{1/6} 
& \mbox{(if $v = \alpha^{1/2}v_{\rm H} $ and $u = \alpha^{-1/12} v_{\rm H}$)},\\
\alpha^{1/2} 
& \mbox{(if $v < \alpha^{1/2}v_{\rm H} $ and $u = v_{\rm H}$)}.\\
\end{array}\right.
 \label{eq:dr1rat}
\end{equation}
Since $\alpha < 1$, the contribution of large bodies is always lower than 
that of small bodies. Equal accretion assumed by \citet{Schlichting2011}
is not justified, as long as Eqs.~(\ref{eq:dr1dtc}) and (\ref{eq:dvdt}) are employed.

It is not necessary to distinguish between the horizontal and vertical velocities in 
Eqs.~(\ref{eq:dr1dtc}) and (\ref{eq:dvdt}) in the dispersion dominant regime ($v/v_{\rm H} > 1$).
However,  their difference plays a crucial role in the shear dispersion regime ($v/v_{\rm H} < 1$).  
The accretion rate in the shear dominant regime only depends on the vertical velocity $v_{\bot}$
\citep[e.g.,][]{Ida1989}. Thus, $v$ is replaced by $v_{\bot}$ if $v < v_{\rm H}$ in  Eqs.~(\ref{eq:dr1dtc}).
The viscous stirring rate of $v_{\bot}$ in the shear dominant regime is smaller by $v_{\bot}/v$ than that given by 
Eq.~(\ref{eq:dvdt}) \citep[their Section~4.4]{Goldreich2004}. This means that if $v < v_{\rm H}$,  
$v_{\bot}$ evolves toward zero as dynamical friction becomes always larger than vertical stirring.
Thus, we may be able to apply the highest possible value for the second term of the r.h.s of Eq.~(\ref{eq:dr1dtc})
as long as $v < v_{\rm H}$ or $u < \alpha^{-1/6} v_{\rm H}$. After taking into account this effect, 
the relative contribution of large bodies turns out as
\begin{equation}
\frac{\mathrm{d}R/\mathrm{d}t|_{\rm large,correct}} {\mathrm{d}R/\mathrm{d}t|_{\rm small}}      = 
\left\{ \begin{array}{ll} 
\alpha^{-1/6}  
& \mbox{(if $u = \alpha^{-1/6} v_{\rm H}$)},\\
1
& \mbox{(if  $u = \alpha^{-1/8} v_{\rm H}$)},\\
\alpha^{1/2} 
& \mbox{(if $u = v_{\rm H}$)}.\\
\end{array}\right.
 \label{eq:dr1rat2}
\end{equation}
In the rage of $ \alpha^{-1/8} < u/v_{\rm H} < \alpha^{-1/6}$,  the large bodies' contribution becomes 
large than the small bodies' contribution.

If growth of large bodies is dominated by mutual merging, 
the velocity dispersion $u/v_{\rm H}$ may be determined by equating
Eq.~(\ref{eq:dudt}) with the second term of the r.h.s of Eq.~(\ref{eq:dr1dtc}) instead of the first term used in Eq.~(\ref{eq:effi}).
This gives 
\begin{equation}
\left(\frac{u}{v_{\rm H}}\right)_{\rm S} = \alpha^{-1/8}, \label{eq:ueqs}
\end{equation}
at which the contributions from large and small bodies are equal (equal accretion). 
The question is whether this state can remain avoiding $u/v_{\rm H}$ from further decrease down to unity  
where the contribution of small bodies dominates.  
We infer that the factors of order of unity 
ignored so far in the stirring and accretion rates are important.   
Let us consider a factor somewhat larger than 
unity in front of the second term of Eq.~(\ref{eq:dr1dtc}) for $v < \alpha^{1/2}v_{\rm H}$.
For simplicity, the same factor is applied to Eq.~(\ref{eq:dudt}). 
The detailed numerical calculations of the accretion rates 
\citep{Ida1989,Greenzweig1990} and the stirring rates \citep{Ida1990}
support this idea.
Then, the big bodies' contribution still dominates even at the equilibrium velocity,
$u = \alpha^{-1/8}v_{\rm H}$, while the small bodies' contribution
takes a fixed value relative to the large bodies'  one.
This continues even at a slightly lower $u$ while
the viscous stirring timescale becomes shorter than the accretion timescale. 
As a result, $u/v_{\rm H}$ increases towards the equilibrium value again. 
We also call this stage after runaway growth, as the trans-Hill stage, although 
$u/v_{\rm H}$ does not exactly go down to unity. 

The planetary mass at the beginning of  the trans-Hill stage with $u/v_{\rm H} = \alpha^{-1/8}$ is given as
\begin{equation}
M = M_{\rm tran}(q) = \left[-\frac{(q+3)^2}{q+2}\right]^{1/(q+2)}\alpha^{3/[4(q+2)]} m_0. \label{eq:mtrans}
\end{equation}
For $q=-2.5$, $M_{\rm tran} = 4 \alpha^{-3/2}m_0$.
The efficiency during the trans-Hill stage is given as 
\begin{equation}
\left(\frac{\Sigma}{\sigma}\right)_{\rm S}  = f_{\rm s} \alpha^{3/4}, \label{eq:sig0s2}
\end{equation}
where $f_{\rm s} $ is the order of unity factor, which we estimate to be 0.5 from our simulations.
This equation is the same as the one derived by \citet{Schlichting2011}. 
Note that, however,
they derived this efficiency assuming equal accretion without justification while
we interpret this efficiency realized as an effect of large-body dominant accretion.
Since the efficiency takes a fixed value independently of $M$, the mass distribution for large bodies 
is expected to have $q = -2$ also in this case (but see the discussion below). 
Inserting Eq.~(\ref{eq:sig0s2}) into Eq.~(\ref{eq:mem}), the planetary mass at the onset of oligarchic growth is  
\begin{equation}
M_{\rm oli,S} = f_{\rm s}^{3/2}  \alpha^{9/8} M_{\rm iso}  \label{eq:molis2}.
\end{equation} 

To summarize, the embryo mass $M_{\rm oli}$ and the velocity dispersion $(u/v_{\rm H})_{\rm oli}$ at the onset of oligarchic growth are
\begin{equation}
M_{\rm oli}  = 
\left\{ \begin{array}{ll} 
M_{\rm oli,S}
& \mbox{(if $ m_0  < m_{\rm 0,crit} $)}, \\ 
M_{\rm oli,L}
& \mbox{(if $m_0  > m_{\rm 0,crit} $)},
\end{array}\right.
 \label{eq:moli}
\end{equation}
and
\begin{equation}
\left(\frac{u}{v_{\rm H}}\right)_{\rm oli}  = 
\left\{ \begin{array}{ll} 
\left(u/v_{\rm H}\right)_{\rm S}
& \mbox{(if $ m_0  < m_{\rm 0,crit} $)}, \\ 
\left(u/v_{\rm H}\right)_{\rm L} 
& \mbox{(if $m_0  > m_{\rm 0,crit} $)},
\end{array}\right.
 \label{eq:uoli}
\end{equation}
where
the critical planetesimal mass $m_{\rm 0,crit}$ that separates these two regimes is given by equating Eq.~(\ref{eq:molis2}) with Eq.~(\ref{eq:molil}) as 
\begin{equation}
m_{\rm 0,crit} = f_{\rm s}^{7/2}\alpha^{21/8} M_{\rm iso}. \label{eq:m0c}
\end{equation}
We use $M_{\rm oli,S}$ as the prediction of the planetary mass at the onset of oligarchic growth for small $m_0$,
although its difference from $M_{\rm oli,SS}$ [Eq.~(\ref{eq:molis})] is quite small. 

The argument here has some unresolved issues.
First, the accretion of large bodies with $v < \alpha^{1/2} v_{\rm H}$
is not neutral but ordered ($\mathrm{d}(\dot{R}/R)/ \mathrm{d}R < 0$).
All large bodies in the ordered regime eventually converge in mass and this leads to 
increase of $\Sigma/\sigma$, whereas we see a constant $\Sigma/\sigma$ or 
$q = -2$ in our simulations for small $m_0$.
Since the mass range of large bodies in the ordered regime is generally very narrow (or $v_{\bot} \sim \alpha^{1/2} v_{\rm H}$) in simulations,
this effect seems limited.
Second,  the solution of $u = v_{\rm H}$ (Section~2.1) is not excluded by the theory. 
The solution of $u = \alpha^{-1/8}v_{\rm H}$ is realized in our simulations probably because 
this solution comes earlier than the one with $u = v_{\rm H}$. 
However,  it is unclear whether the former one is preferentially realized in arbitrary initial conditions
such as one starting with $u = v_{\rm H}$ and $\Sigma/\sigma \sim \alpha$.

We have discussed the contribution of large bodies during the trans-Hill stage.
To complete the section, we discuss their contribution during the runway and oligarchic growth stages.
The large bodies' contribution during oligarchic growth is estimated to be half of the small bodies' contribution if $b_e$ is fixed
\citep{Chambers2006}. The large bodies' contribution can be larger if $b_e$ increases with time \citep{Morishima2013}.
If mutual merging between protoplanets is prohibited in oligarchic growth,  their influences of gravity eventually overlap ($b_e$ decreases)
and this violates the definition of oligarchic growth.
Therefore, equal accretion or large-body dominant accretion is actually a 
necessary condition for the onset of oligarchic growth, consistent with the correction given in this subsection.

The argument for  the large bodies' contribution during runway growth is complicated.
For $v > v_{\rm H}$,  the theory predicts that the contribution of small bodies is dominant  [Eq.~(\ref{eq:dr1rat})].
However, numerical simulations \citep[see also Section~3]{Ormel2010b}
show that the contribution of large bodies is comparable to that of small bodies even during the runway growth stage
and different interpretations than those for the trans-Hill stage are necessary.
We attribute the discrepancy to an overestimation of $v$ in the analytic model.
If Eq.~(\ref{eq:dvdt}) is applied to a continuous size distribution, one can find that
$v' \propto m^{-1/4}$ for $M_{\rm stir} < m < M$, where $v'$ is the velocity dispersion of bodies with mass $m$, 
while $v'$ is constant ($= u$) for  $m < M_{\rm stir}$ \citep{Goldreich2004}.
On the other hand, simulations of runaway growth show $v' \propto m^{-1/4}$ 
down to the smallest bodies while the mass distribution has $q \simeq -2.5$ \citep{Morishima2008}.
It is likely that the velocity distribution does not fully evolve to the equilibrium state as the growth timescale of large bodies is shorter 
\citep[see also][]{Rafikov2003}. 
Taking this effect into account, one can find that large and small bodies contribute 
almost equally to growth of large bodies during runaway growth 
($\mathrm{d} \dot{M}/\mathrm{d}\log m_{\rm imp} \propto m_{\rm imp}^{q+2-2\beta}  \sim {\rm const}$, 
where $m_{\rm imp}$ is the impactor's mass and $\beta = \mathrm{d} \log v'/ \mathrm{d} \log m_{\rm imp}$, 
provided that all bodies are in the dispersion dominant regime). 
The combination of the values of $\beta \simeq -1/4$ and $q \simeq -2.5$ 
is also consistent with the steady-state mass distribution, $q = -13/6+\beta$ \citep{Makino1998}.
The argument here does not modify $M_{\rm oli,L}$ [Eq.~(\ref{eq:molil})], 
as the derivation was simply based on the assumption of $q = -2.5$.

\section{Numerical simulations}
\subsection{Simulation codes}

We simulate planetary accretion that starts from equal-mass planetesimals each with the mass of $m_0$, 
using two different codes. 
If $m_0$ is large and the initial number of planetesimals $N$ is not very large, 
we perform pure $N$-body simulations using the parallel-tree code, pkdgrav2 \citep{Morishima2010}.
The gravitational solver employed by the code is a parallelized tree method, 
in which the gravitational forces from distant particles are calculated using the multipole expansions 
\citep{Richardson2000,Stadel2001}. The computational cost for the gravity calculation is proportional to 
only $N \log N$ with tree methods.
The code employs the symplectic integrator, SyMBA \citep{Duncan1998}, that handles 
close encounters between planetesimals using multi time-stepping. 
The combination of the advanced gravity solver and the integrator allows us to handle 
a large number of particles while keeping a large time step. 
Unfortunately, pure $N$-body simulations are still computationally intense and cannot practically handle
$N$ more than $\sim 10^4$, provided that a number of time steps is 
hundreds of millions for a simulation of oligarchic growth.

To overcome such a situation, we recently have developed a particle-based hybrid code \citep{Morishima2015}.
We use this new code if $m_0$ is small and $N$ is very large.
The code retains various advantages of direct $N$-body calculations as the gravitational accelerations 
due to planetary embryos are calculated by the $N$-body routine.
The code can also handle a large number of small planetesimals using the super-particle approximation, 
in which a large number of small planetesimals are represented by a small number of tracers.
Collisional and gravitational interactions between tracers are handled by a statistical routine that uses 
the phase-averaged collision and stirring rates.
As planetesimals grow, the number of planetesimals in a tracer decreases. Once the number of planetesimals in a tracer
becomes unity, this particle is promoted to a sub-embryo. After the sub-embryo becomes massive enough, 
it is further promoted to a full-embryo.  The acceleration of the sub-embryo due to surrounding planetesimals 
is handled by the statistical routine to avoid artificially strong kicks on the sub-embryo while
the acceleration of the full-embryo is always handled by the $N$-body routine.
The algorithms and various tests for the validation of the code are described in detail in \citet{Morishima2015}. 

In a hybrid simulation, the tracer mass $m_{t0}$, which is inversely proportional to the initial number of tracers $N_{\rm tr}$,
needs to be chosen. 
Since $m_{t0}$ is also a minimum mass of a sub-embryo handled by the code, 
it is desirable that $m_{t0}$ is lower than $M_{\rm oli}$ [Eq.~(\ref{eq:moli})] to examine the onset of oligarchic growth in detail.
This condition is roughly fulfilled in all the simulations (Table~1).

\subsection{Initial conditions and collisional outcome}

\begin{table}
\begin{center}
\footnotesize

\begin{tabular}{|cccc|} 

\hline
Region (AU)                        &  0.09-0.11           & 0.9-1.1 & 27-33        \\
$f_{\sigma}$                       & 1                &1   & 10            \\ 
$M_{\rm total}$ (g) & $6.3 \times 10^{26}$           & $2.0 \times 10^{27}$  &   $4.6 \times 10^{29}$         \\ 
$M_{\rm iso}$ (g)   & $1.0 \times 10^{26}$          &  $5.8 \times 10^{26}$  &    $2.0 \times 10^{30}$         \\ 
$\delta t$ (days)                 & 0.2                        & 6                & 3600         \\
$m_0$ (Hybrid) (g)                  & \multicolumn{3}{c|}{$10^{13}$, $10^{17}$, $10^{21}$} \\
$m_0$ ($N$-body) (g)              & $3.1\times 10^{23}$&$10^{24}$ & N/A \\     
$m_{t0}$ (g)                               & $6.3 \times 10^{22}$ & $2.0 \times 10^{23}$ & $4.6 \times 10^{25}$ \\
$M_{\rm oli,S}$  (g)                    & $1.5 \times 10^{24}$ & $6.4 \times 10^{23}$ & $4.9 \times 10^{25}$ \\  \hline                         
\end{tabular}

\end{center}

Table~1. Parameters used for simulations. 
Region represents the initial inner and outer edges of the planetesimal annulus.  
$f_{\sigma}$ is the scaling factor for the solid surface density relative to MMSN.
$M_{\rm total}$ is the initial total mass of planetesimals.
$M_{\rm iso}$ is the isolation mass for $b_{\rm e} = 10$[Eq.~(\ref{eq:miso})].
$\delta t$ is the time step for orbital integration. 
$m_0$ is the initial mass of a planetesimal.
$m_{t0}$ is the initial mass of a tracer for hybrid simulations with $N_{\rm tr} = 10^4$.
$M_{\rm oli,S}$ is the predicted embryo mass at the onset of oligarchic growth for small $m_0$ [Eq.~(\ref{eq:molis2})]. 
\end{table}

The initial surface density of planetesimals is given by 
\begin{equation}
\sigma = f_{\sigma}\sigma_{\rm M1}\left(\frac{a}{\rm 1\hspace{0.3em} AU}\right)^{-1.5},
\end{equation}
where $f_{\sigma}$ is the scaling factor relative to the Minimum Mass Solar Nebula (MMSN) \citep{Hayashi1981} and 
$\sigma_{\rm M1}$ is the surface density scaled at 1 AU  given as 
\begin{equation}
\sigma_{\rm M1} = 
\left\{ \begin{array}{ll} 
7.1 {\rm \hspace{0.3em} g \hspace{0.3em} cm}^{-2}
& \mbox{(if $a < $ 2.7 AU)}, \\ 
30  {\rm \hspace{0.3em} g \hspace{0.3em} cm}^{-2}
& \mbox{(if $a > $ 2.7 AU)}.
\end{array}\right.
\end{equation}
Since it is computationally too intense 
to perform simulations of the entire disk,  we choose annuli of planetesimals at three different regions: $a = 0.1$, 1, and 30 AU.
The width of each annulus is set to be 0.2$a$, so the effect of radial diffusion is unimportant as long as 
$e < 0.1$, where $e$ is the orbital eccentricity.
We adopt $f_{\sigma} = 1$ for the suites of simulations at $0.1$ and 1 AU while a large surface density, $f_{\sigma} = 10$, is adopted 
for the simulations at 30 AU to reduce accretion time. The various parameters for each suite of simulations are summarized in Table~1.

The internal density of any bodies is assumed to be 2 g cm$^{-3}$.
We adopt three initial masses of planetesimals: $m_0 =$ $10^{13}$, $10^{17}$, and $10^{21}$ g ($s = 0.1, 2.3$, and 49 km) for all three regions.
For these simulations, we use our hybrid code. 
The initial number of tracers for hybrid simulations is $N_{\rm tr} = 10000$. 
Simulations with larger $N_{\rm tr}$ are computationally too intense since many embryos form near the onset of oligarchic growth.
To check the resolution effect we also perform the same simulations staring with $N_{\rm tr} = 1000$ for some cases. 
In addition to the hybrid simulations, we perform pure $N$-body simulations with $N =$ 2000.
This number gives $m_0 = 3.1 \times 10^{23}$ g  and $10^{24}$ g ($s = 335$ km and 492 km) for $a =0.1$ and 1 AU, respectively.
We do not perform pure $N$-body simulations at 30 AU since practically no collision occurs within the age of the solar system 
for such a small $N$.

The initial velocity dispersion $u$ is set to be the escape velocity of an individual planetesimal and the initial ratio of 
the rms eccentricity $\langle e \rangle^{1/2}$ to the rms inclination $\langle i \rangle^{1/2}$ is 2.
Similar results are obtained even with a lower initial $u$ because $u$ quickly increases due to viscous stirring.  
We do not consider any effects of a gaseous disk, such as gas drag, for comparison with our analytic theory.
Simulations are performed at least up to the midst of oligarchic growth and each run takes typically a few cpu weeks.
Some of simulations are performed twice for the same input parameters but using different random numbers 
for creation of the initial positions and velocities. Since the statistical variations in mass and velocity evolutions are found to be 
small enough, we will show results of only one simulation for each set of input parameters.



For the outcome of collisions, we adopt the merging criterion proposed by \citet{Genda2012}.
They showed that two colliding bodies merge if the impact velocity $v_{\rm imp}$ is lower than the critical velocity $v_{\rm cr}$,
while a collision with $v_{\rm imp} > v_{\rm cr}$ results in an inelastic rebound.
If $v_{\rm imp}$ is very large,  collisional fragmentation occurs in reality
\citep{Benz1999,Leinhardt2012},
although we ignore this effect in our simulations as well as our analytic theory.
If perfect accretion is applied, on the other hand, the masses of the smallest planetesimals unnaturally 
increase through mutual collisions even after their impact velocities become much larger than the mutual escape velocities.
Such an effect is again not considered in the analytic theory.

Consider an impact between the target with mass $m_i$ and the impactor with mass $m_j$ ($\le m_i$).
Let the impact angle be $\theta_{\rm c}$ ($\theta_{\rm c} = 0$ for a head-on collision). 
Based on thousands of SPH simulations, 
\citet{Genda2012} derived the formula of $v_{\rm cr}$ as 
\begin{equation}
\frac{v_{\rm cr}}{v_{\rm esc}} = c_1 \Gamma \Theta^{c_5} + c_2 \Gamma + c_3 \Theta^{c_5} + c_4, \label{eq:gen}
\end{equation}
where $v_{\rm esc}$ is the mutual escape velocity, $\Gamma = (m_i- m_j)/(m_i + m_j)$, and $\Theta = 1-\sin{\theta_{\rm c}}$. 
The coefficients are $c_1 = 2.43$, $c_2 = -0.0408$, $c_3 = 1.86$, $c_4 = 1.08$, and $c_5 = 5/2$.
While we use Eq.~(\ref{eq:gen}), our simulation results are not  sensitive to choice of $v_{\rm cr}/v_{\rm esc}$ as long as it is somewhat larger than unity. 
The impact velocity vector is decomposed to the normal and tangential components ($v_n$ and $v_t$)
relative to the vector pointing toward the center of the target from the center of the impactor. 
If the impact is judged as a hit-and-run collision,
the tangential component  
$v_{t} $ is assumed to be unchanged and
the normal component of the post-impact relative velocity $v_n'$ is given as 
\begin{equation}
v_n' = 
\left\{ \begin{array}{ll} 
0
& \mbox{(for $v_t  > v_{\rm esc} $)}, \\ 
-\left(v_{\rm esc}^2-v_t^2\right)^{1/2}
& \mbox{(otherwise)}.
\end{array}\right.
\label{eq:vnd}
\end{equation}
This velocity change is similar to but slightly different
from that adopted by \citet{Kokubo2010}, who always set $v_n' = 0$
and adjust the post-impact $v_t$.
We do not consider any mass exchange
between the target and the impactor for a hit-and-run collision.
Appendix~E of \citet{Morishima2015} describes in detail how hit-and-run collisions are handled in the hybrid code.

\subsection{Results}

\begin{figure*}
\begin{center}
\includegraphics[width=0.9\textwidth]{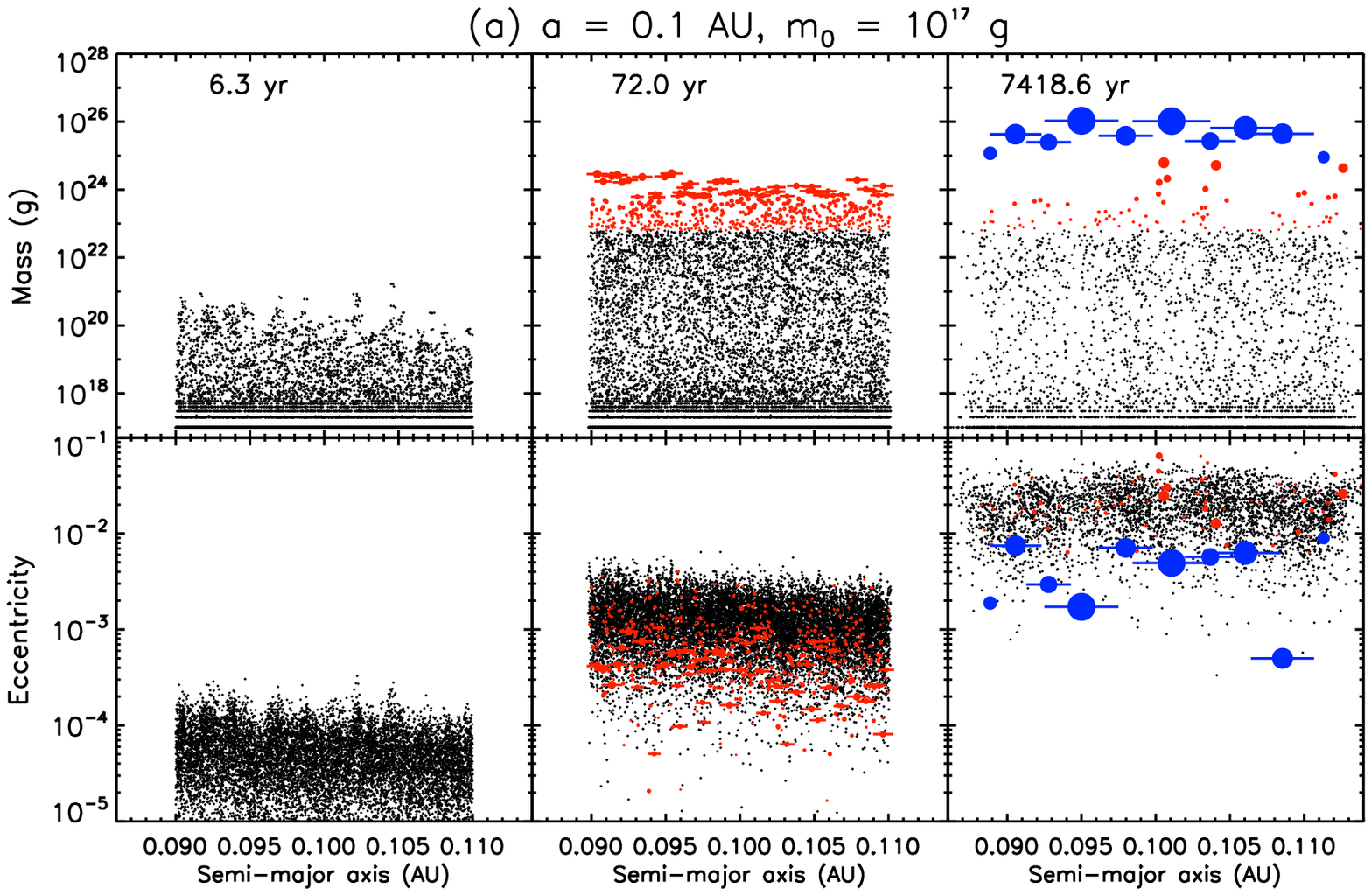}
\includegraphics[width=0.9\textwidth]{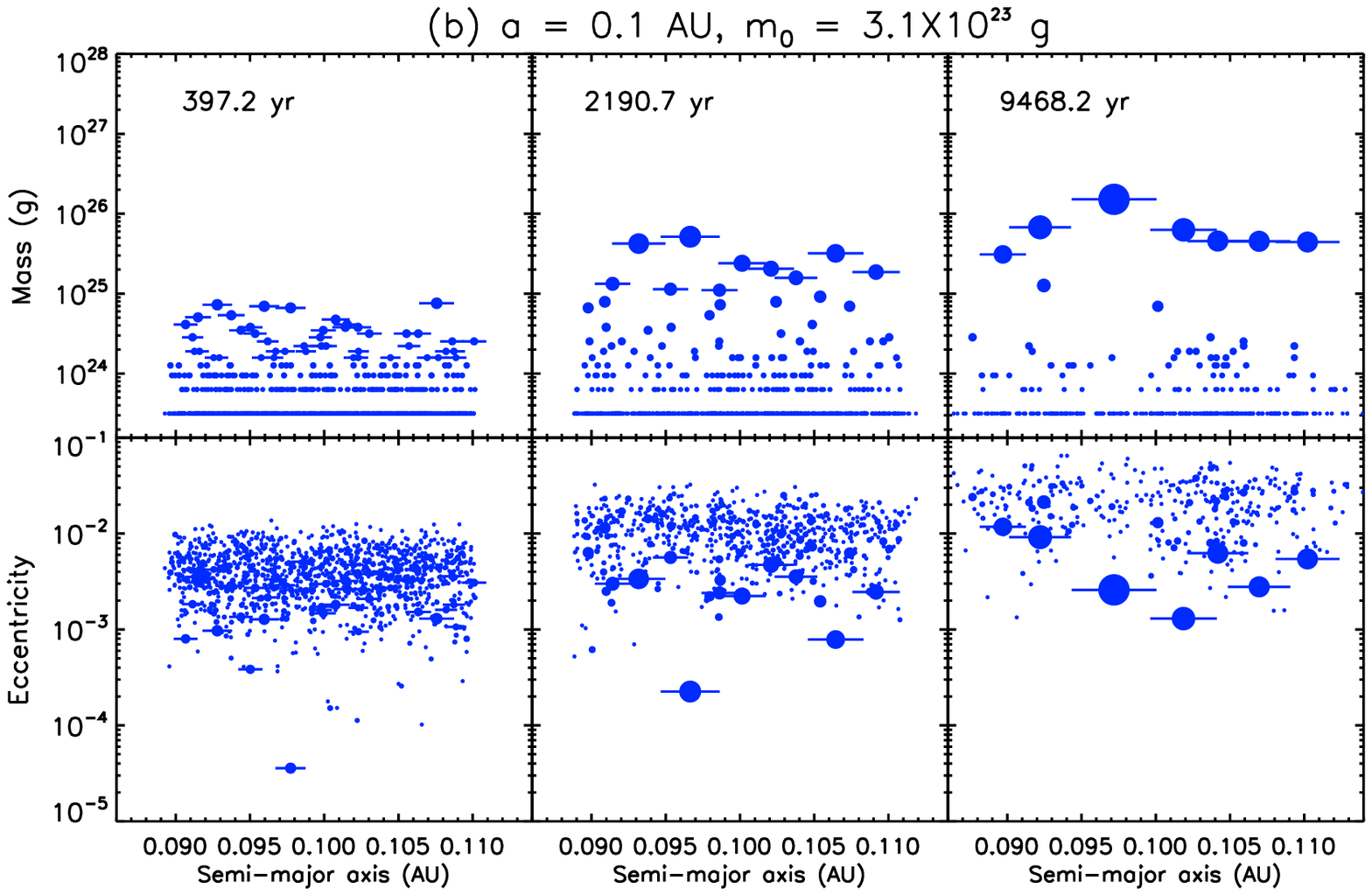}
\end{center}
\caption{Snapshots of mass and eccentricity distributions for four selected simulations. 
For hybrid simulations (a,c,d), tracers are black dots, sub-embryos ($m_{t0} \le m < 100 m_{t0}$) and full embryos ($ m \ge 100 m_{t0}$) 
are red and blue circles, respectively. The circle's radius for an embryo is proportional to its physical radius. 
For tracers, plotted masses are those of planetesimals in them, not tracers masses.
The horizontal bars are given for embryos more massive than 0.2$M$ and the half length of the bar is 10 Hill radii
of each embryo.}
\label{fig:avsem}

\end{figure*}

\begin{figure*}
\begin{center}
\includegraphics[width=0.9\textwidth]{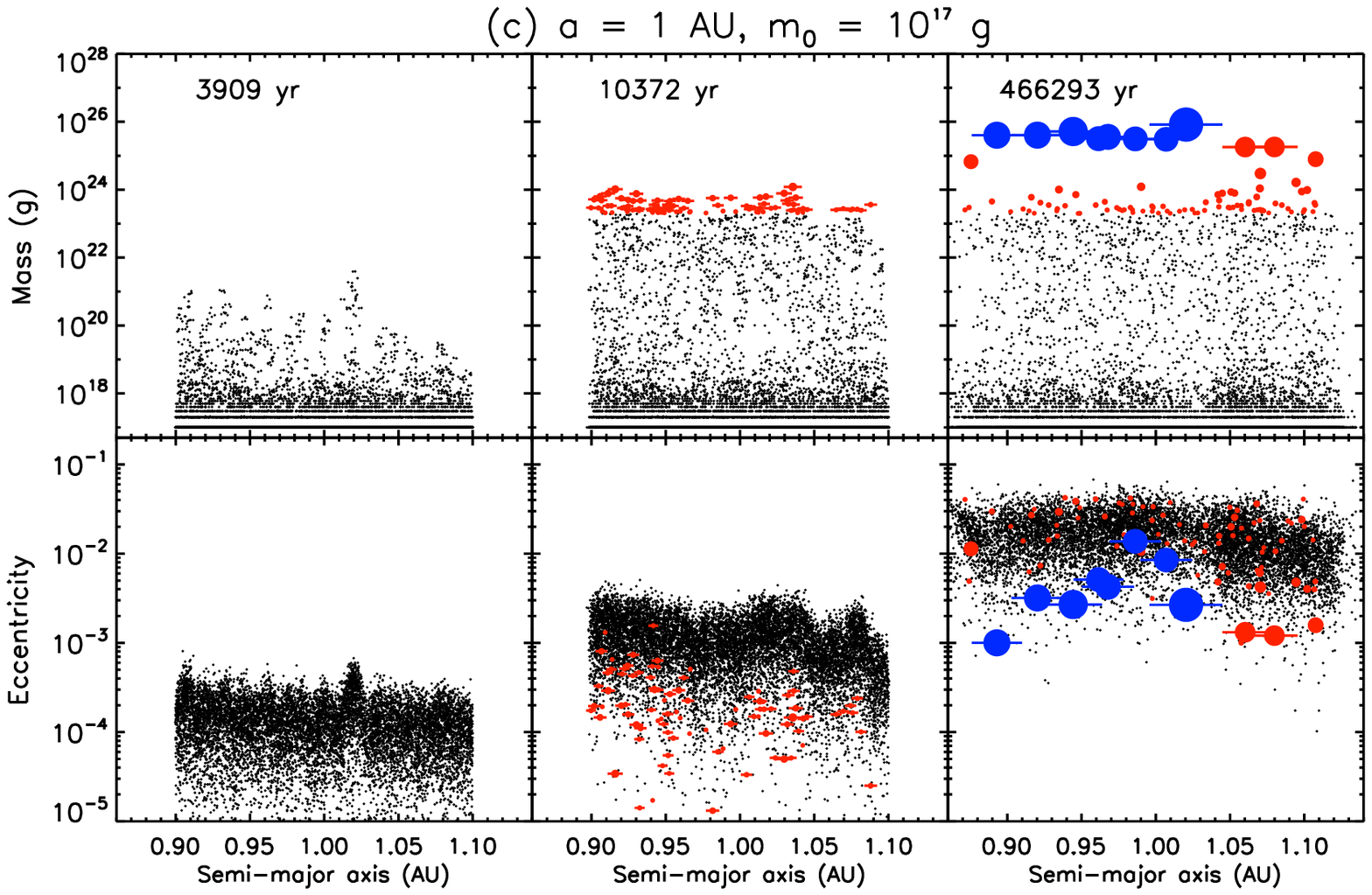}
\includegraphics[width=0.9\textwidth]{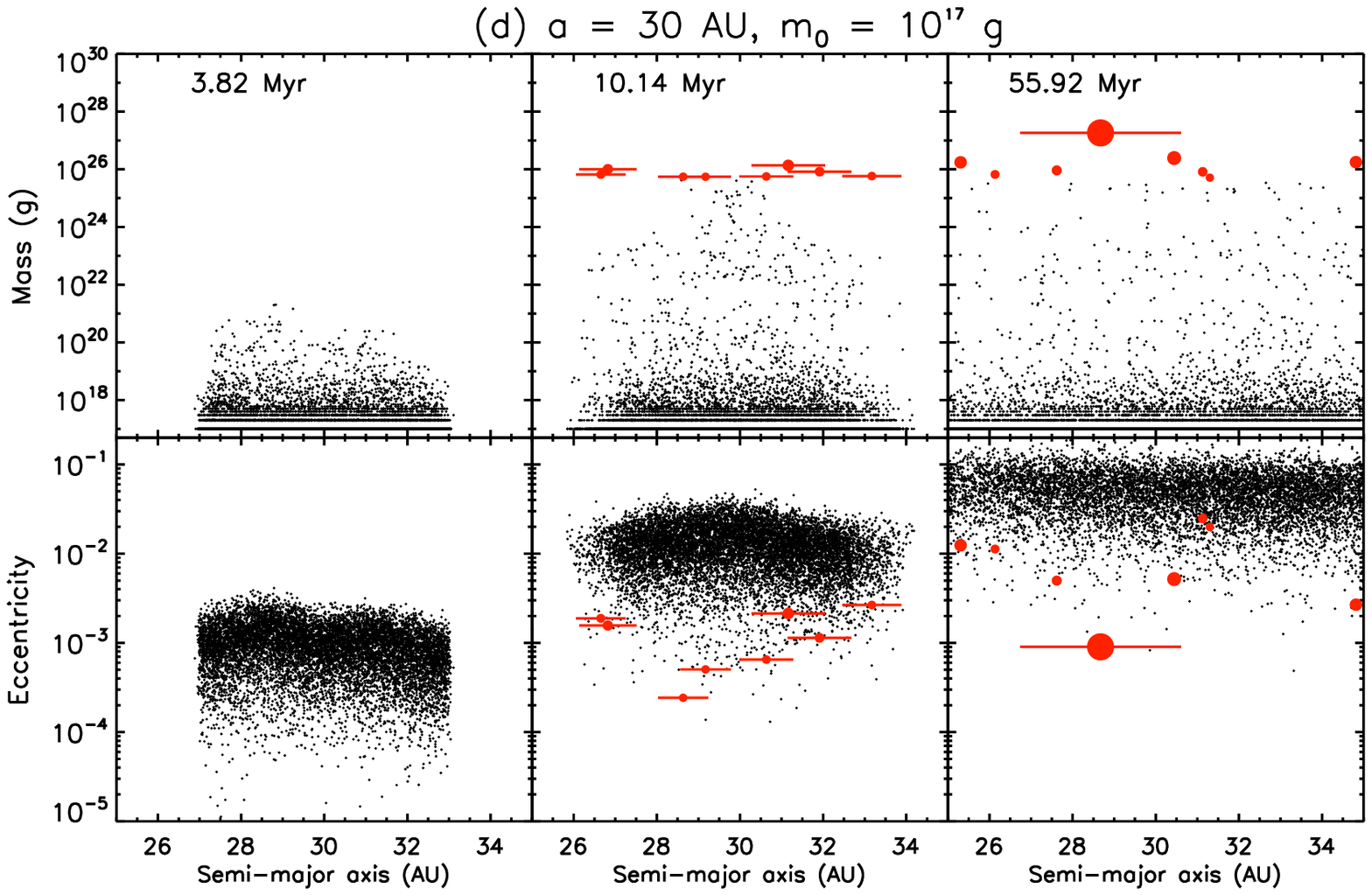}
\end{center}
\nonumber
Figure~\ref{fig:avsem} - continue.
\end{figure*}

\begin{figure*}
\begin{center}
\includegraphics[width=0.6\textwidth]{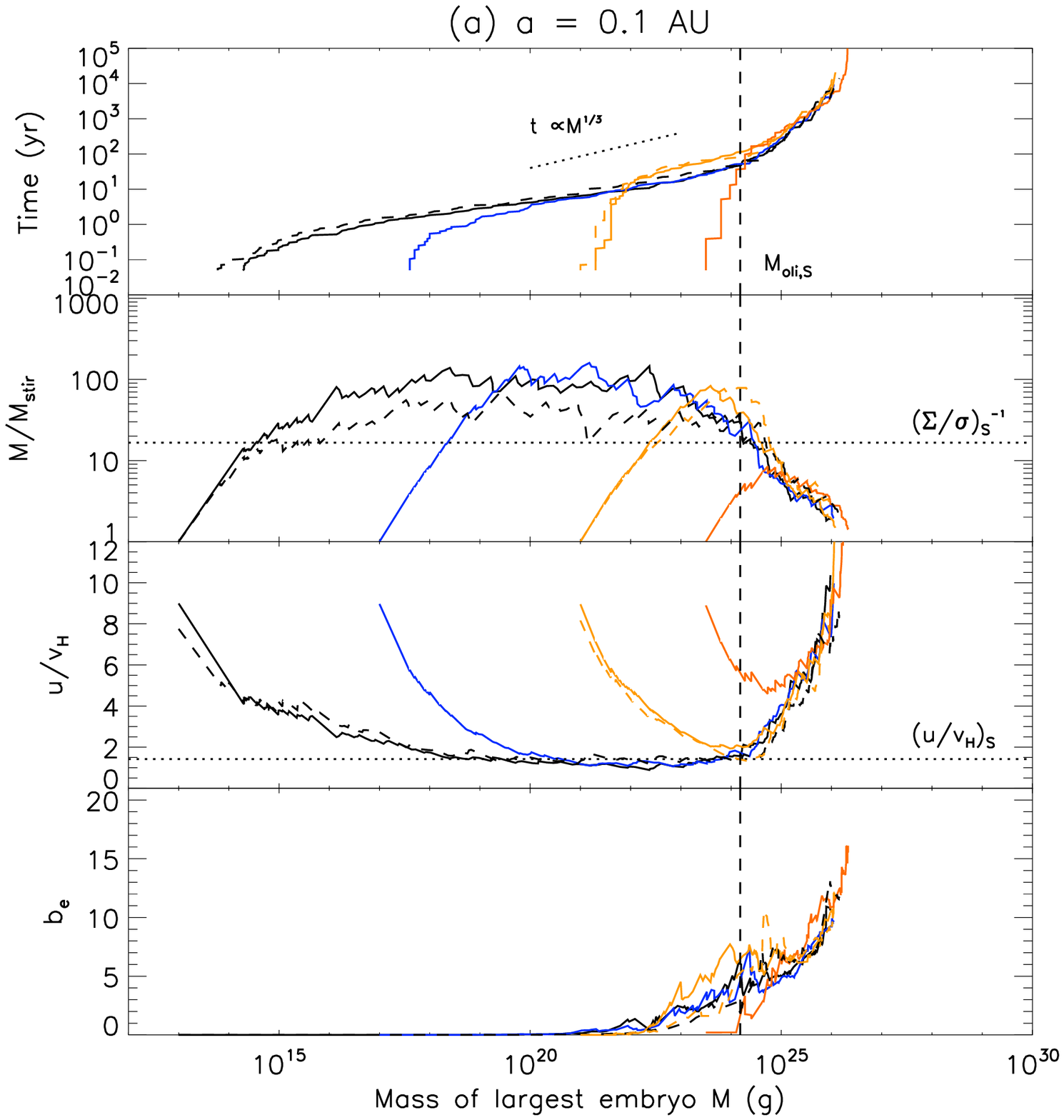}
\includegraphics[width=0.6\textwidth]{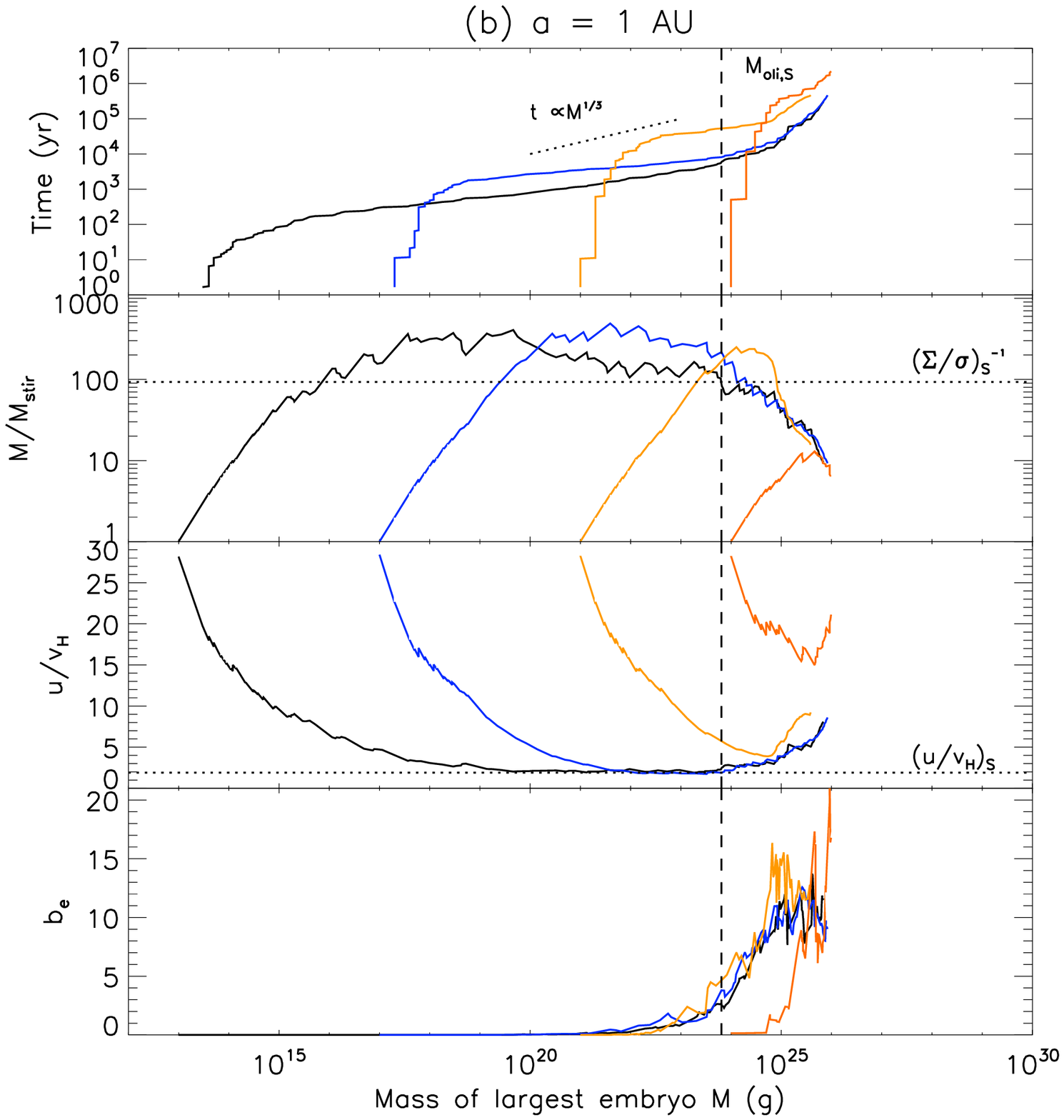}
\end{center}
\caption{Time $t$, $M/M_{\rm stir}$, $u/v_{\rm H}$, and $b_{\rm e}$
as a function of largest body's mass $M$ for 
$a = $ 0.1,  1,  and 30 AU.
For hybrid simulations ($m_0 = 10^{13}, 10^{17}$, and $10^{21}$ g), 
$N_{\rm tr} = 10000$ (solid) and 1000 (dashed). The vertical dashed line is $M_{\rm oli,S}$ [Eq.~(\ref{eq:molis2})].
The slope for $t$ and the values for $M/M_{\rm stir} (= (\Sigma/\sigma)_{\rm S}^{-1})$ and $u/v_{\rm H}$ 
expected from the theory for small $m_0$ are shown by the dotted lines. }
\label{fig:vsm}
\end{figure*}

\begin{figure*}
\begin{center}
\includegraphics[width=0.6\textwidth]{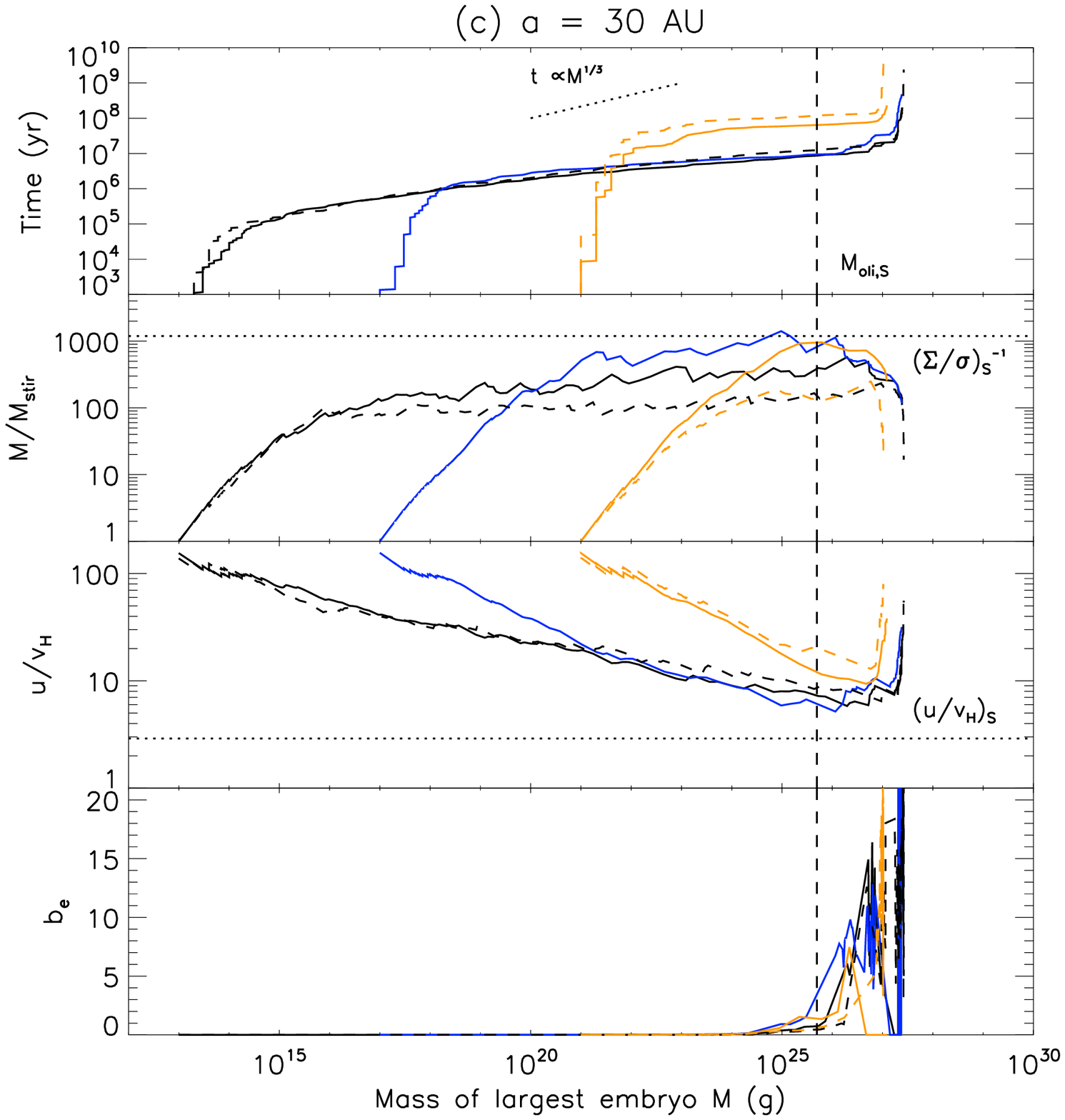}
\end{center}
Figure~\ref{fig:vsm}-continue.
\end{figure*}

The overall results well agree with the theoretical predictions described in Section~2. 
We briefly recap how oligarchic growth commences for 
different input parameters.  Simulations at $a = $ 30 AU slightly suffer from resolution effects 
and one needs to see the results with some caution. We will discuss these effects in the next section.

Figure~\ref{fig:avsem} 
demonstrates how planetary accretion proceeds in our simulations. 
In the early stage, some large planetesimals grow rapidly and they locally exert velocity dispersion.
Eventually embryos form and they grow with keeping some orbital separations, as typical for 
oligarchic growth.  
For simulations at $a = 0.1$ and 1 AU, about ten oligarchic bodies 
form in the end while only one distinctively large embryo form for the run at $a = 30$ AU. 
Late time planetary accretion at $a = 30$ AU is halted primarily by
depletion of the local surface density of planetesimals due to their radial diffusion,
although its effect is insignificant around the onset of oligarchic growth where $M \sim M_{\rm oli}$.

To see the time evolution of each simulation more quantitatively, we plot in Fig.~\ref{fig:vsm}
the time $t$, the largest bodies mass $M$ relative to the effective mass 
$M_{\rm stir} (= \langle m^2 \rangle/\langle m \rangle)$, the normalized velocity $u/v_{\rm H}$, 
and the normalized orbital separation of large bodies $b_e$
taking $M$ in the horizontal axis. 
We define the velocity $u/v_{\rm H}$ as 
\begin{equation}
\frac{u}{v_{\rm H}} \equiv \frac{\langle e^2 \rangle^{1/2}_{0} + \langle i^2 \rangle^{1/2}_{0}}{h_{M}},
\end{equation}
where $\langle e^2 \rangle^{1/2}_0$ and $\langle i^2 \rangle^{1/2}_0$ are the rms eccentricity and inclination 
of smallest bodies $(m = m_0)$
and $h_{M}$ is the reduced Hill radius of the largest body. 
Statistical fluctuation of $h_{M}$ is likely to be small since
the mean value for the five largest bodies is found to be very close to $h_{M}$.
We measure the normalized orbital separation $b_{\rm e}  $ as
\begin{equation}
b_{\rm e} = \frac{1}{N_{\rm L} -1}\sum_i^{N_{\rm L} -1} b_{ij}, \hspace{1em}
b_{ij} = \frac{a_j - a_i}{(k_i + k_j)(a_i+a_j)h_{ij}'/4}, \label{eq:be}
\end{equation}
where $N_{\rm L}$ is the number of particles (embryos or tracers) with $m > 0.2M$,
the index $i$ is in ascending order of $a$, $j = i +1$, and
$k_{i}$ is the number of bodies in a tracer and unity for an embryo.
We approximately applied the following form of the reduced mutual Hill radius $h_{ij}'$ only for measuring 
$b_{\rm e}$ 
\begin{equation}
h_{ij}' = \left(\frac{m_{ij}}{3M_{\odot}}\right)^{1/3}, \hspace{1em} m_{ij}' = 2\frac{k_im_i +k_jm_j}{k_i +k_j}.
\end{equation}
This form is reduced to the standard reduced mutual Hill radius if both neighboring bodies are embryos ($k_i = k_j = 1$).
In Eq.~(\ref{eq:be}), bodies less massive than $0.2M$ are not counted, since they are scattered by large bodies and behave like small planetesimals. 
Neighboring pairs with $b_{ij} > 30$ are not counted since they are likely to grow in spatially different regions with
little gravitational interactions.

The very beginning of planetary accretion takes place 
by collisions between small planetesimals which have comparable collisional and geometric cross sections.
Once $M$ becomes more massive than $\sim 10 m_0$, runaway growth starts to occur due to gravitational focusing. 
With increasing $M$, $M/M_{\rm stir}$ increases and $u/v_{\rm H}$ decreases. 
This accelerates the growth of the largest body. 

If $m_0$ is small, $u/v_{\rm H}$ eventually reaches as low as unity and 
$M/M_{\rm stir}$ takes its highest value which is comparable to the theoretically expected value.
This low  $u/v_{\rm H}$  state, or the trans-Hill stage, continues until $M$ reaches $M_{\rm oli,S}$. 
The mass increases as $M \propto t^{3}$ ($dR/dt =$ const) as long as 
$u/v_{\rm H}$ is nearly fixed during this stage.
Some cases have steeper dependence since $u/v_{\rm H}$ 
gradually decreases with $M$.
As the largest body grows more massive than $M_{\rm oli,S}$, $u/v_{\rm H}$ starts to increase.
This is because now the accretion timescale of the largest body is longer than the stirring timescale.  
The orbital separation of large bodies normalized by their mutual Hill radius at this time is about 3-10.

If $m_0$ is not small, $u/v_{\rm H}$ starts to increase without decrease of $u/v_{\rm H}$ down to near unity. 
In this case, the planetary mass at this turnaround point is about $M_{\rm oli,L}$. 
Similar trends are seen regardless of $a$.

\begin{figure*}
\begin{center}
\includegraphics[width=0.7\textwidth]{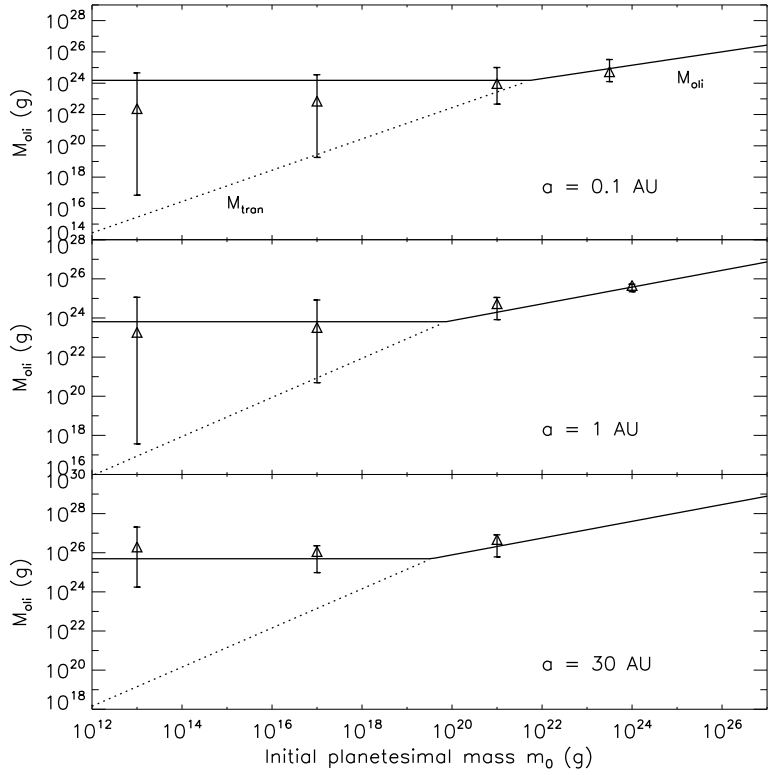}

\end{center}

\caption{
Planetary mass $M_{\rm oli}$ at the onset of oligarchic growth as a function of initial planetesimal mass $m_0$
for $a = $ 0.1,  1,  and 30 AU. The values from numerical simulations are shown by triangles with error bars 
(see the text how we derived them)
while the theoretical prediction [Eq.~(\ref{eq:moli})] is shown by a solid line in each panel.
The dotted line is $M_{\rm tran}$ [Eq.~(\ref{eq:mtrans})] for $q = -2.5$. }
\label{fig:moli1}

\end{figure*}

\begin{figure*}
\begin{center}
\includegraphics[width=0.7\textwidth]{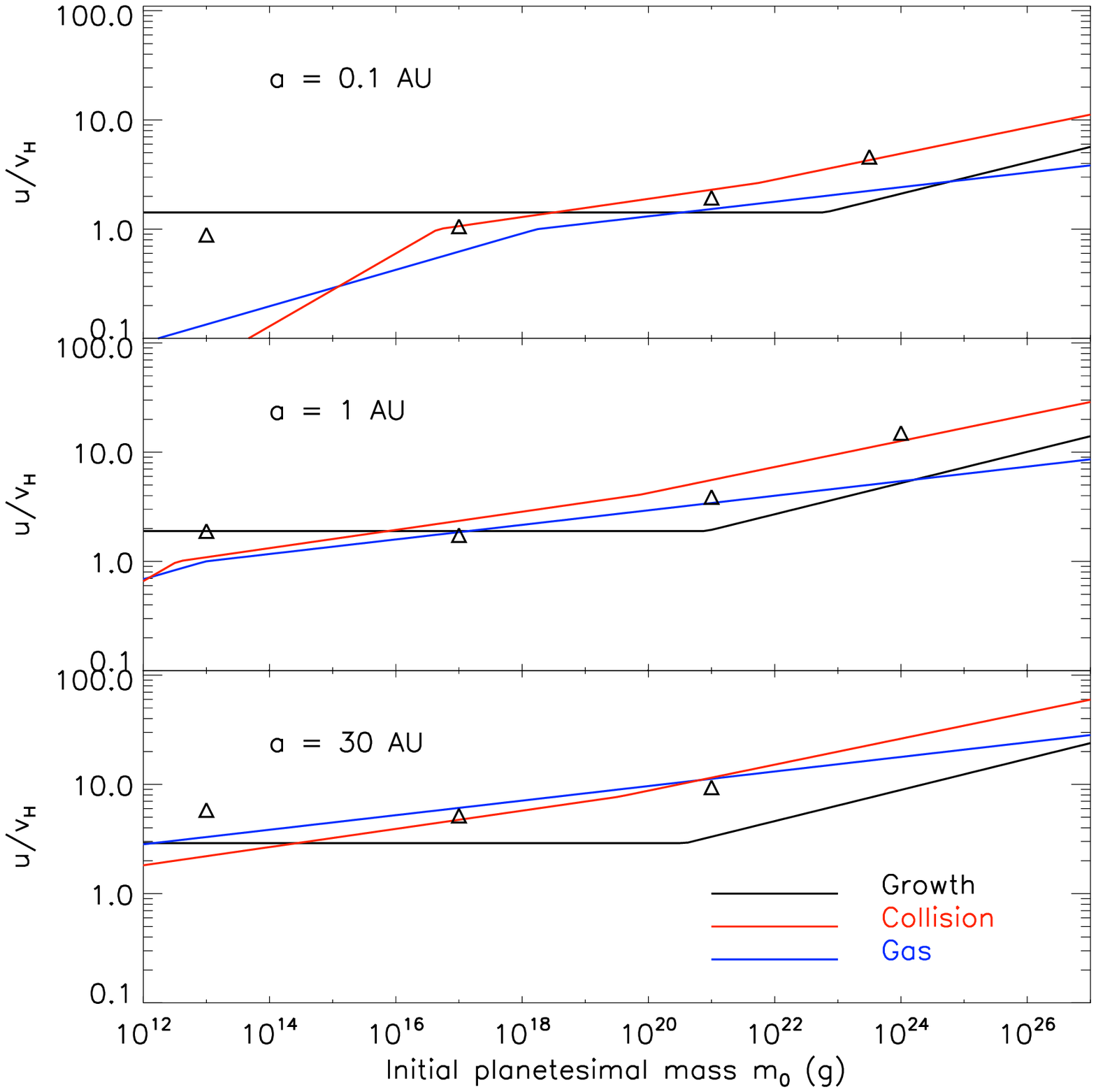}
\end{center}
\caption{Velocity dispersion at the onset of oligarchic growth or minimum $u/u_{\rm H}$.
Triangles are the values from simulations while black curves are theoretical predictions [Eq.~(\ref{eq:uoli})]
in the case without any damping force. The equilibrium velocities determined by balance 
between viscous stirring and collisional damping are shown by red curves.
The equilibrium velocities determined by balance 
between viscous stirring and gas drag  are shown by blue curves. See Section~4.5 for 
the red and blue curves.
}
\label{fig:damp}
\end{figure*}

In the same manner of \citet{Ormel2010a}, we regard $M$ at the minimum 
$u/v_{\rm H}$ as the planetary mass $M_{\rm oli}$ at the onset of oligarchic growth.
This mass for each simulation is shown as a triangle in Fig.~\ref{fig:moli1}, and the minimum $u/v_{\rm H}$ 
is plotted in Fig.~\ref{fig:damp}.
If $m_0$ is small, this mass may not be very meaningful since 
$u/v_{\rm H}$ fluctuates around a similar value for a wide range of $M$.
Hence we also measure the low velocity range in the $M$-space in which 
$u/v_{\rm H} < {\rm MIN}(u/v_{\rm H})  + \Delta $.
The offset  $\Delta $ should be as small as possible but large enough compared with 
the fluctuation amplitude in $u/v_{\rm H}$.
We choose $\Delta =  (u/v_{\rm H})_{\rm S} = \alpha^{-1/8}$, and 
the range is shown by the error bars for each simulation in Fig.~\ref{fig:moli1}.
This range can be roughly regarded as the trans-Hill stage, although 
the width of the range in the $M$-space is overestimated due to a finite $\Delta $.
We find that $M_{\rm oli}$ and the minimum $u/v_{\rm H}$ from the numerical simulation 
reasonably agree with the theoretical prediction.
Even if the agreement of $M_{\rm oli}$ is not very good for some simulations with low $m_0$,
the range of $M$ in the low velocity stage well encompasses the theoretically predicted $M_{\rm oli}$.
For low $m_0$, the trans-Hill stage begins when $M$ reaches $\sim M_{\rm tran}$ [Eq.~(\ref{eq:mtrans})] as predicted, except 
for the cases of $a = 30$ AU.

\begin{figure*}
\begin{center}
\includegraphics[width=0.45\textwidth]{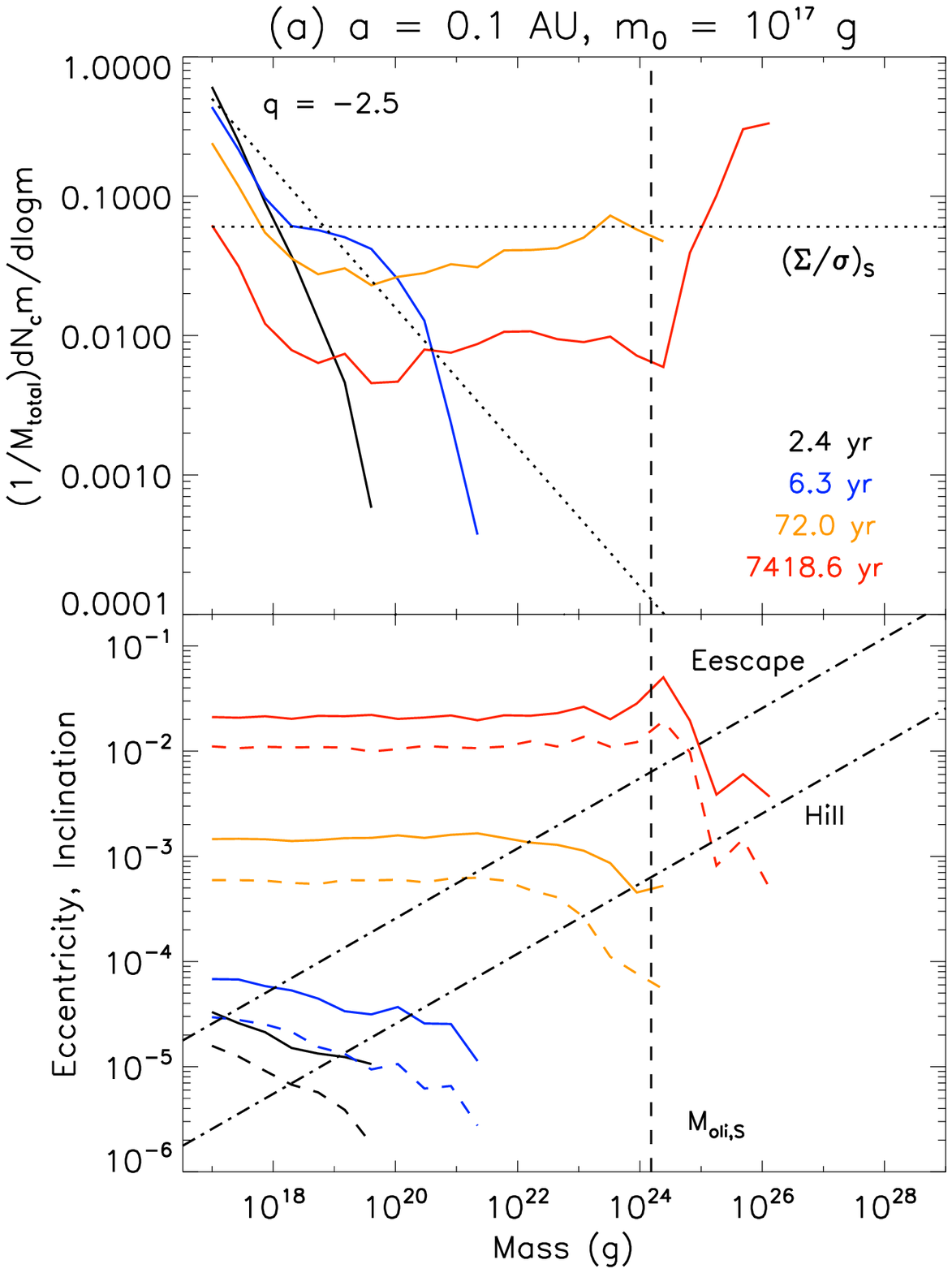}
\includegraphics[width=0.45\textwidth]{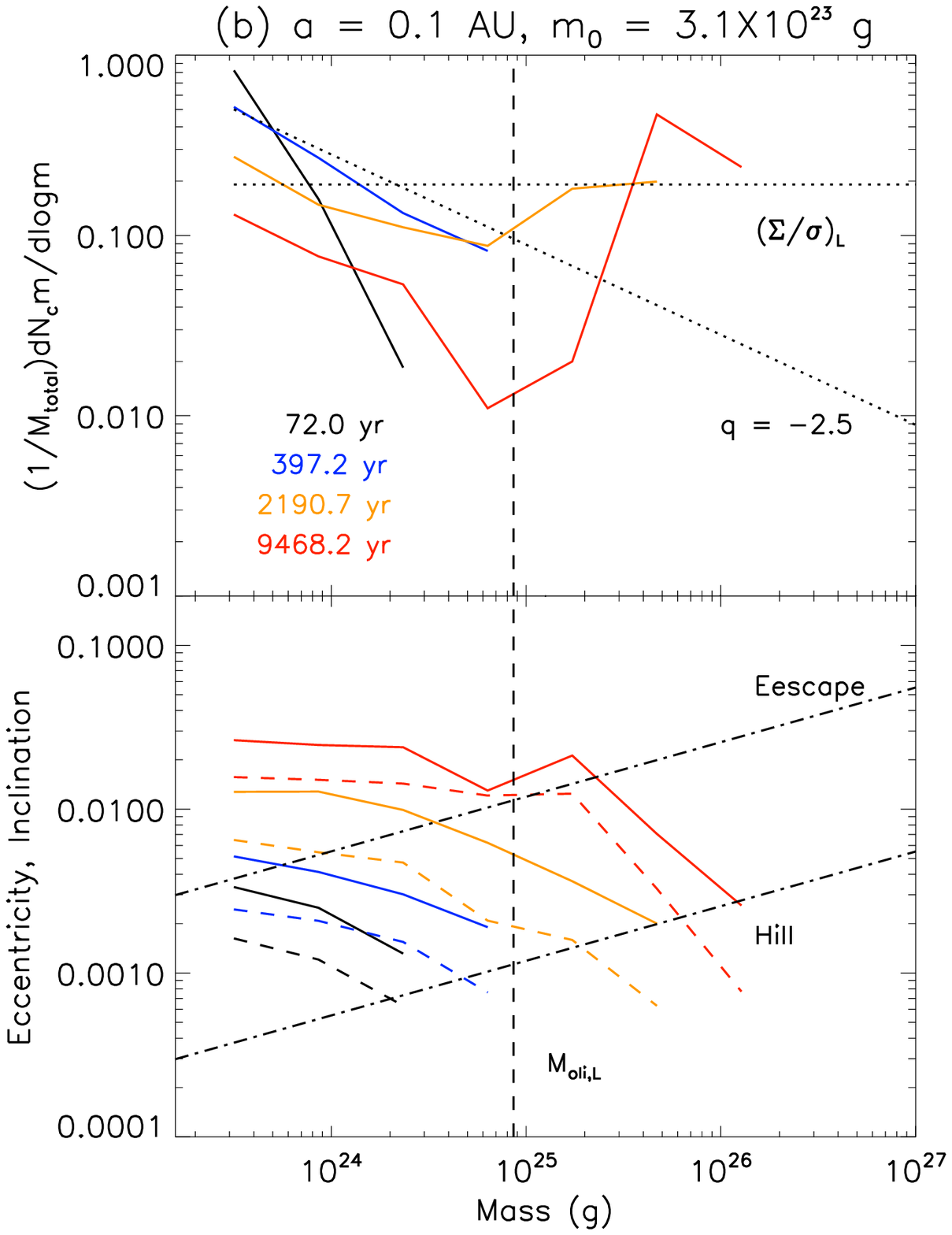}
\includegraphics[width=0.45\textwidth]{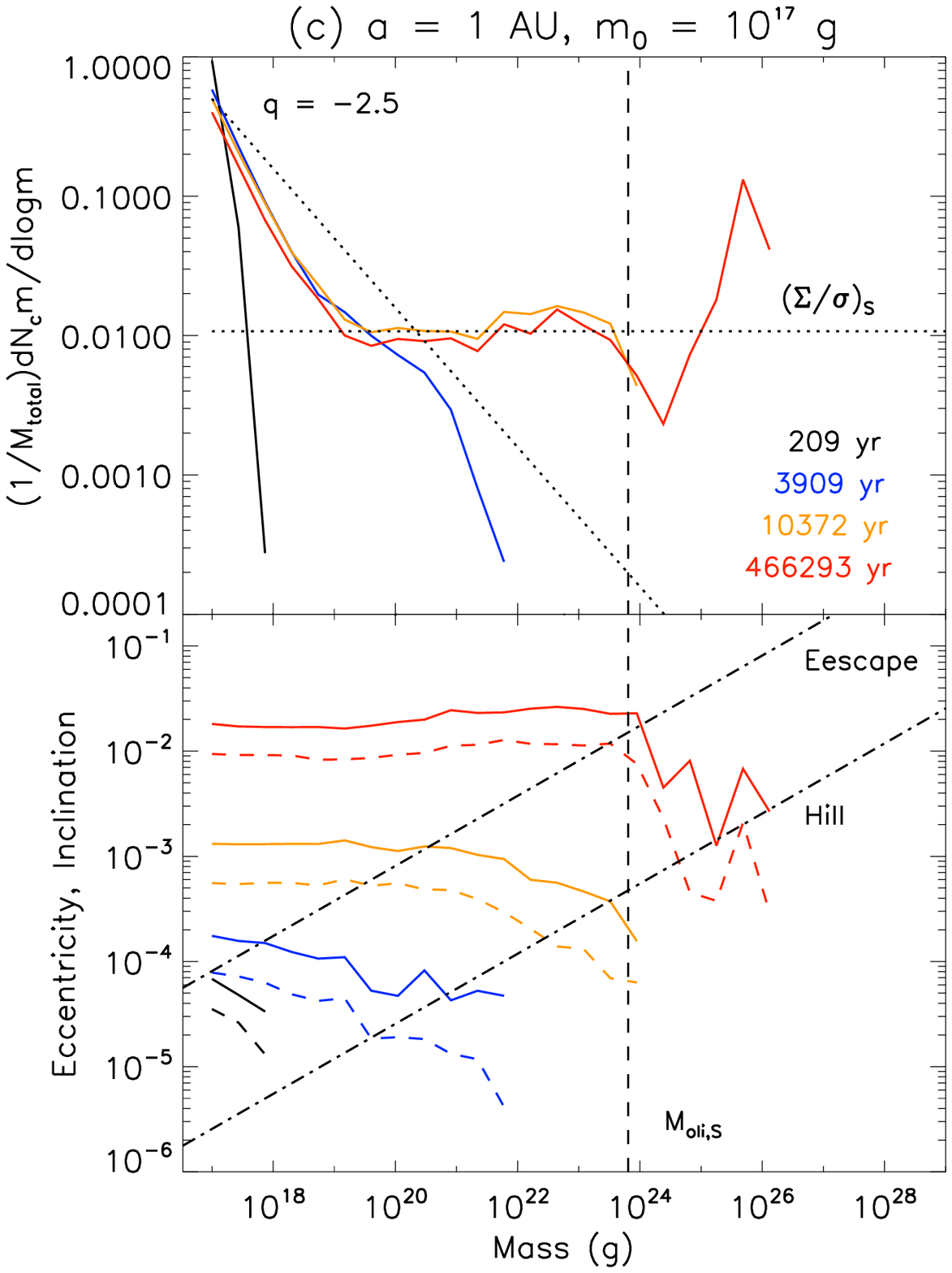}
\includegraphics[width=0.45\textwidth]{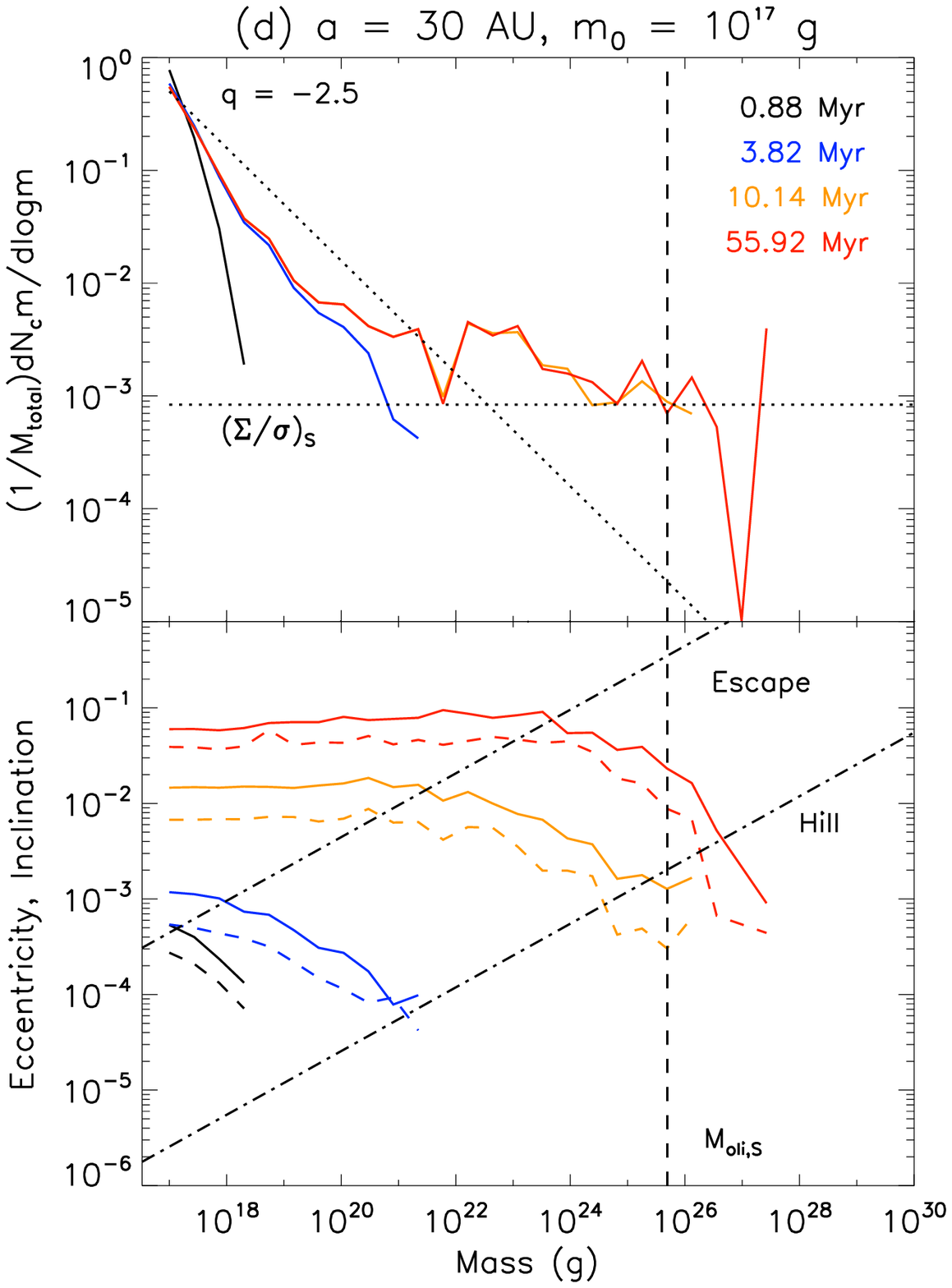}
\end{center}

\vspace{-1.5em}
\caption{
Snapshots of the mass and velocity distributions for four simulations selected in Fig.~\ref{fig:avsem}.
The vertical dashed line is the predicted $M_{\rm oli}$.
In the top panel, the flat dotted lines is the predicted $\Sigma/\sigma$ at the onset of oligarchic growth
and the inclined dotted line shows a slope of $q = -2.5$. 
In the bottom panel, the colored curves are the eccentricity (solid) and inclination (dashed) distributions and 
the two dot-dash lines correspond to the escape and hill velocities at that mass.
}
\label{fig:mdf}
\end{figure*}

\begin{figure*}
\begin{center}
\includegraphics[width=0.46\textwidth]{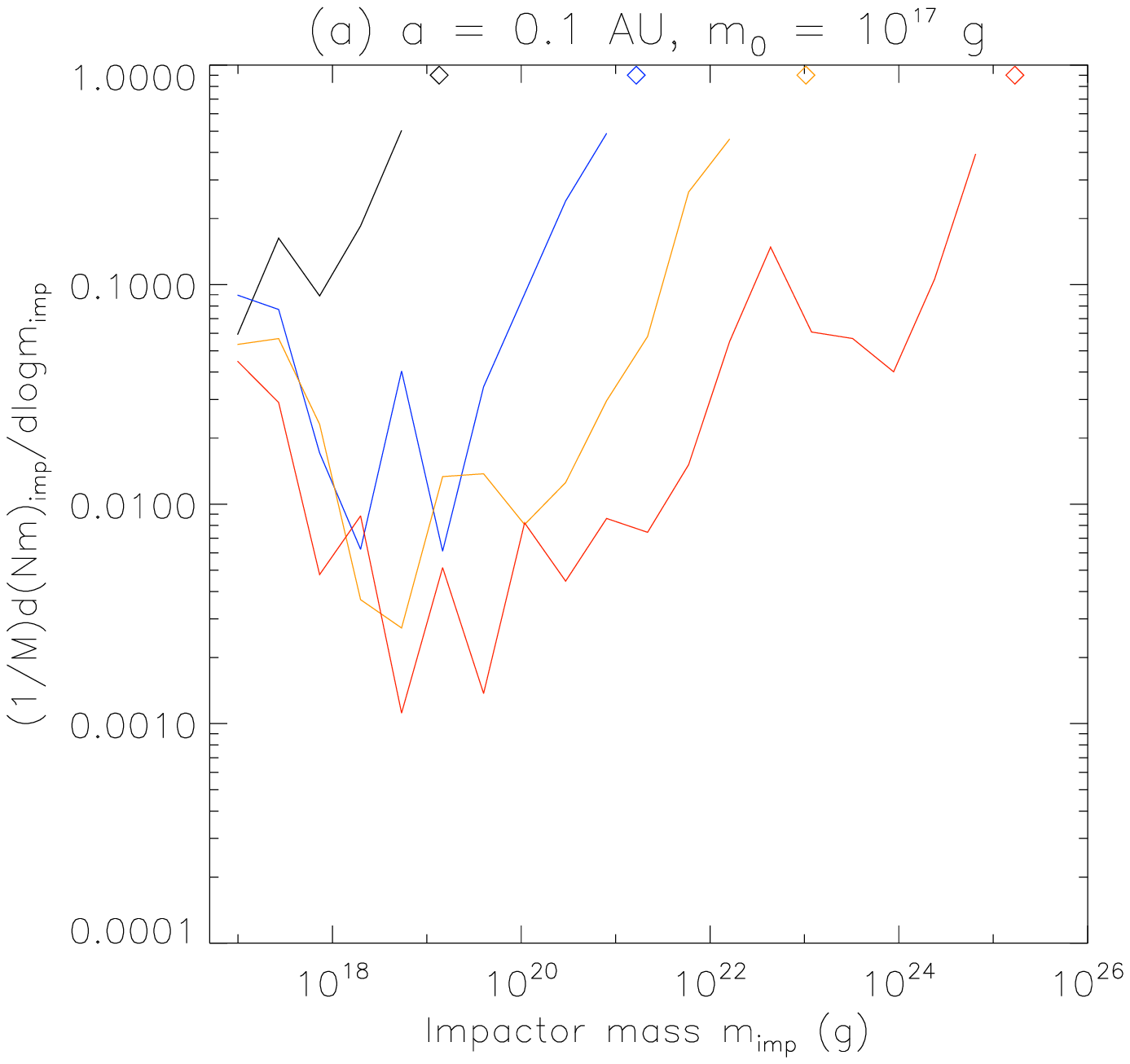}
\includegraphics[width=0.46\textwidth]{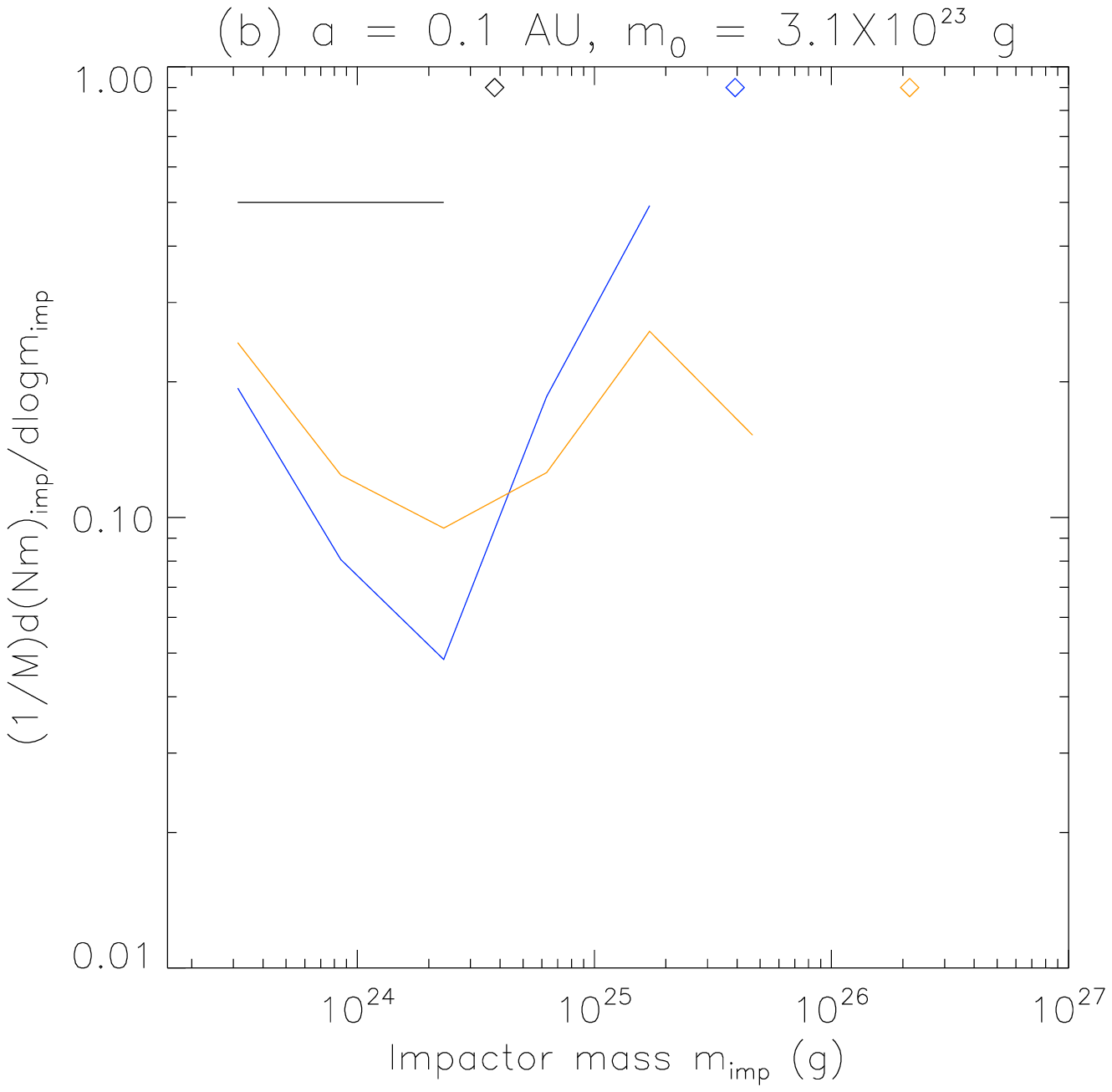}
\includegraphics[width=0.46\textwidth]{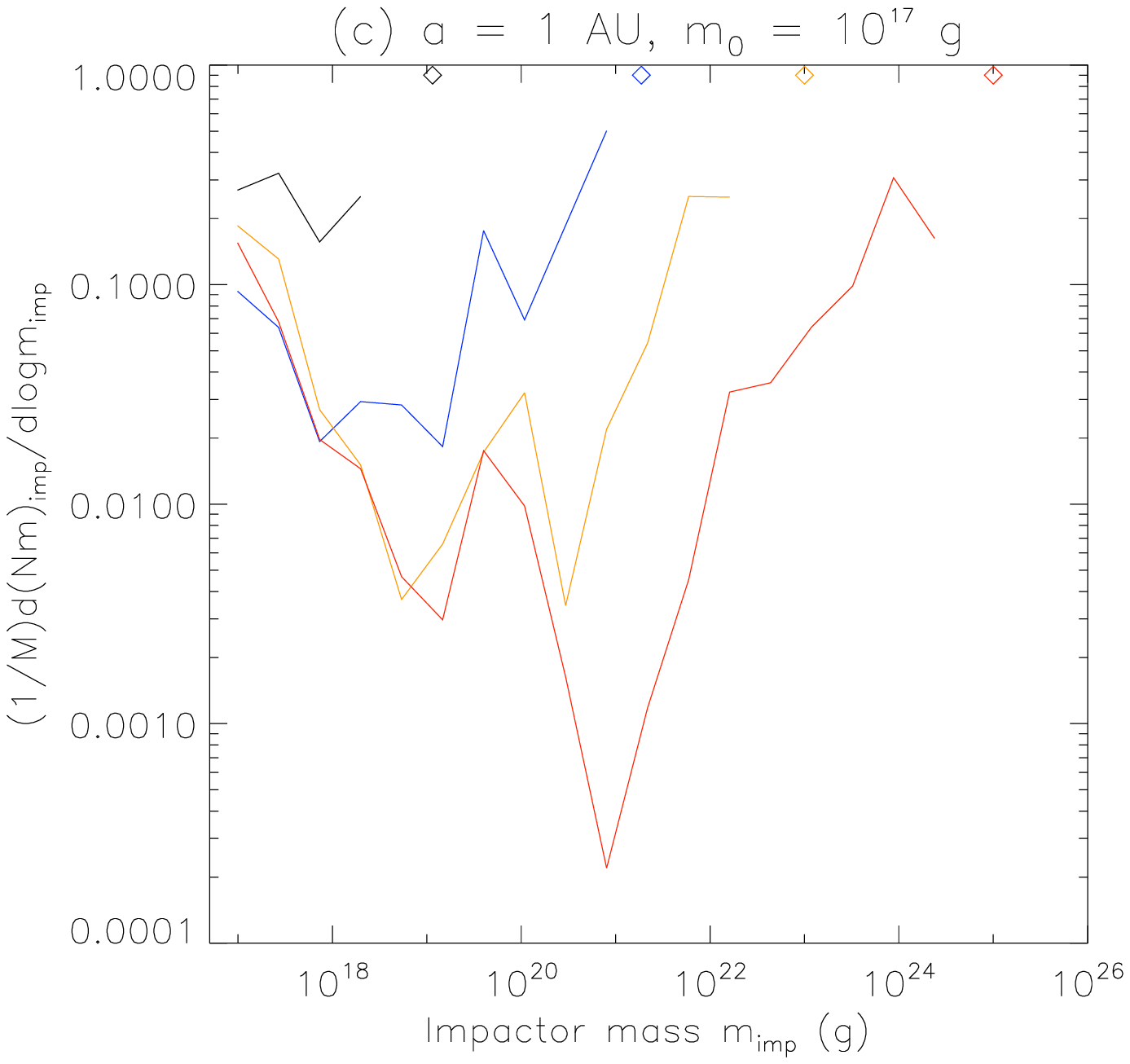}
\includegraphics[width=0.46\textwidth]{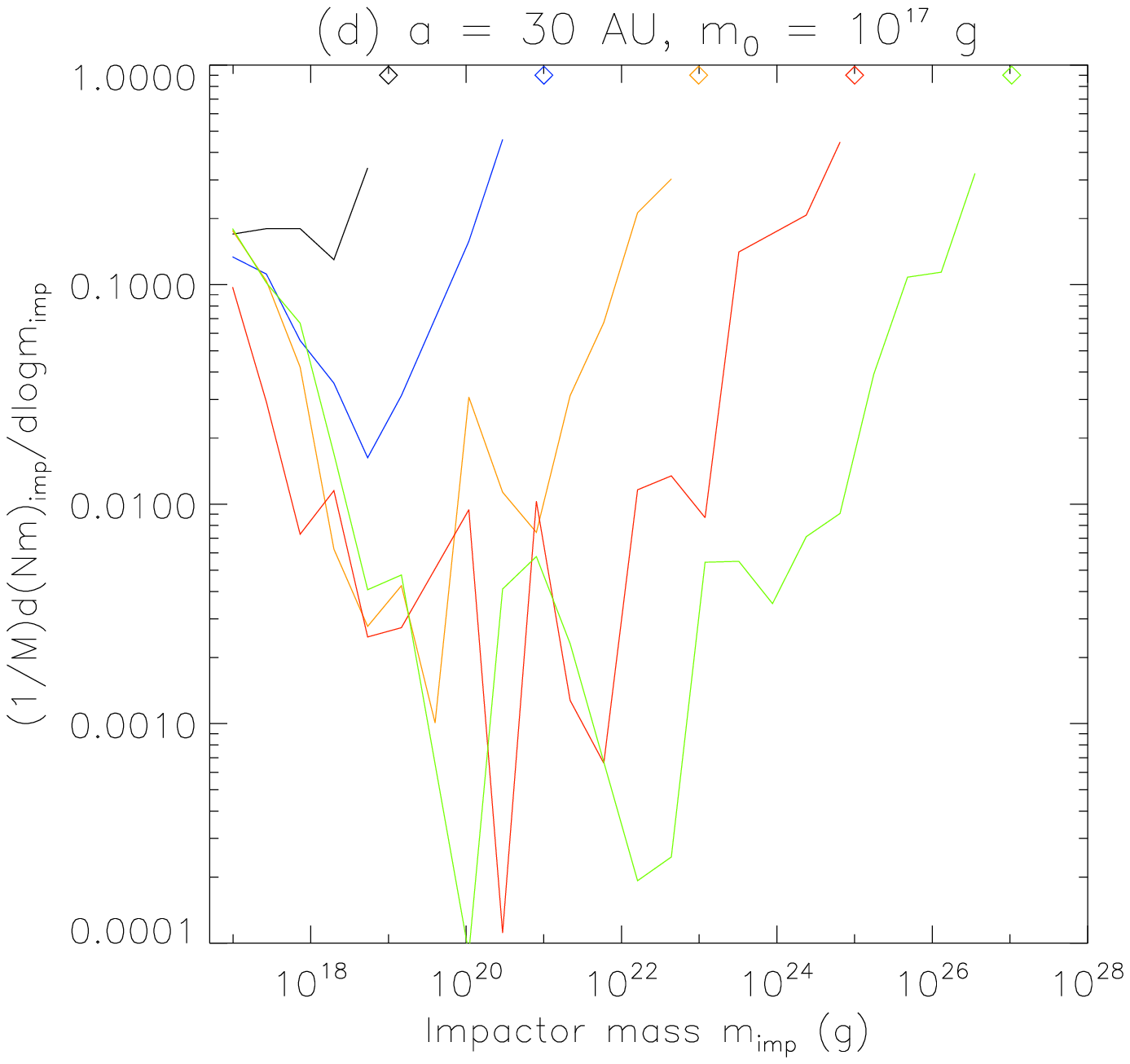}
\end{center}

\caption{Mass distribution of impactors that collided with the largest body until a certain time
for the same simulations selected in Fig.~\ref{fig:avsem}. 
The largest body is selected by its mass at the end of each simulation.
The largest body's mass at each time is shown by a diamond and the corresponding 
impactor's distribution is shown by a solid curve with the same color of the diamond.}
\label{fig:imdf}

\end{figure*}

Figure~\ref{fig:mdf} shows snapshots of the mass and velocity distributions for selected simulations. 
During runaway growth, $q \simeq -2.7$ to $-2.5$, 
as reported in the literature \citep{Kokubo1996,Barnes2009,Ormel2010a}.
During the trans-Hill stage that appears in simulations with small $m_0$, 
$q$ for the massive bodies is about $-2.0$ (a flat line in the top panel).
The mass fraction contained in the most massive bodies, or the efficiency  
$\Sigma/\sigma$ at the onset of oligarchic growth well agrees with the theoretical prediction.
The dependence of $\Sigma/\sigma$ on $\alpha$ for small $m_0$ is 
weaker than $\propto \alpha$ predicted by Eq.~(\ref{eq:sig0s}) but close to $\propto \alpha^{3/4}$  given by Eq.~(\ref{eq:sig0s2}).
During the oligarchic growth stage, oligarchic bodies efficiently merge together 
while the mass distribution slope of planetesimals ($m < M_{\rm oli}$) retains.

Due to the balance between viscous stirring and dynamical friction, 
the rms eccentricity and inclination distributions follow $e, i \propto m^{-1/4}$ 
during the runaway and trans-Hill stages
for bodies with their escape velocities larger than the velocity dispersion (see the last paragraph of Section~2.2).
The velocity is nearly flat for smaller bodies. 
In the very early stage of runway growth ($q < -3$), 
the energy equipartitioning is nearly achieved ($e, i  \propto m^{-1/2}$)
since the heating effect of dynamical 
friction (ignored in Section~2.2) is more important than viscous stirring of large bodies. 
For the same reason,
the energy equipartitioning is generally achieved for oligarchic bodies,
as their mutual stirring is inefficient.

Figure~\ref{fig:imdf} shows the mass distribution of impactors that merge with the largest body.
We find that the contribution of merging with similar-sized bodies 
is comparable to or somewhat larger than the contribution of small bodies 
in all the simulations at any time. During the trans-Hill stage, 
efficient merging between large bodies is realized as they have very low inclinations
due to inefficient vertical stirring (see Fig.~\ref{fig:mdf}). 
A similar result is found in the simulation of \citet{Ormel2010b}.

\subsection{Resolution effects of hybrid simulations}

For the simulations at $a = 0.1$ AU, $M_{\rm oli}$ derived from numerical simulations is almost in convergence, 
as we see good agreements in $u/v_{\rm H}$ between simulations 
with $N_{\rm tr} = 1000$ and $10000$ (Fig.~\ref{fig:vsm}a). This is expected since $m_{t0} < M_{\rm oli,S}$
for both simulations. We also performed simulations with $N_{\rm tr} = 10^5$ for some cases only up to midst of the trans-Hill stage.
We found that $u/v_{\rm H}$ at its minimum is slightly lower than that for $N_{\rm tr} = 10000$
and this may cast doubt on numerical convergence. 
However, the mass spectrum falls off very steeply ($q < -3$) in the most massive tail  for $N_{\rm tr} = 10^5$
so the largest bodies do not significantly contribute to stirring. 
If we interpret $h_{M}$ as the effective reduced Hill velocity of massive bodies 
that contribute to stirring the most for consistency with the theory, $u/v_{\rm H}$  at its minimum 
is already in good convergence for $a = 0.1$ AU.

The simulations at $a = 30$ AU, on the other hand, do not reach convergence,
as $M_{\rm oli}$ and $u/v_{\rm H}$  at its minimum significantly decrease with increasing $N_{\rm tr}$ (Fig.~\ref{fig:vsm}c). 
This is also expected since $m_{t0} > M_{\rm oli,S}$ for $N_{\rm tr} < 10000$.
However, we expect that the simulation with $N_{\rm tr} = 10000$ is 
near convergence, as it gives $M_{\rm oli}$ only slightly larger than the prediction.

This argument seems to be supported by other studies.
The coagulation simulation of \citet{Ormel2010b} at $a = 35$ AU with $s = 1$ km ($m_0 \simeq 10^{16}$ g) (their Fig.~16)
shows that $u/v_{\rm H}$  reaches its minimum value, $\simeq 3$, at $M \sim10^{23}$ g and the trans-Hill stage 
continues until $u/v_{\rm H}$  starts to increase around $M \sim 10^{24}$ g. 
This mass is quite close to $M_{\rm oli,S}$ that we predict for $\sigma$ and $m_0$ $(< m_{\rm 0,crit})$ that they employed.
The minimum value of $u/v_{\rm H}$  is about 5 in our simulations 
for $m_0 < m_{\rm 0,crit} $ (Fig.~\ref{fig:vsm}c) and is close to their value.
The simulation of \citet{Shannon2015} at $a = 45$ AU with $s= 1$ km also shows a result similar to \citet{Ormel2010b}.
The efficiency $\Sigma/\sigma$ of $\sim 10^{-3}$ found by \citet{Schlichting2011} and 
\citet{Shannon2015} is also close to ours (Fig.~\ref{fig:mdf}d).

\section{Discussion}

\subsection{Planetary accretion overview}

\begin{figure*}
\begin{center}
\includegraphics[width=0.7\textwidth]{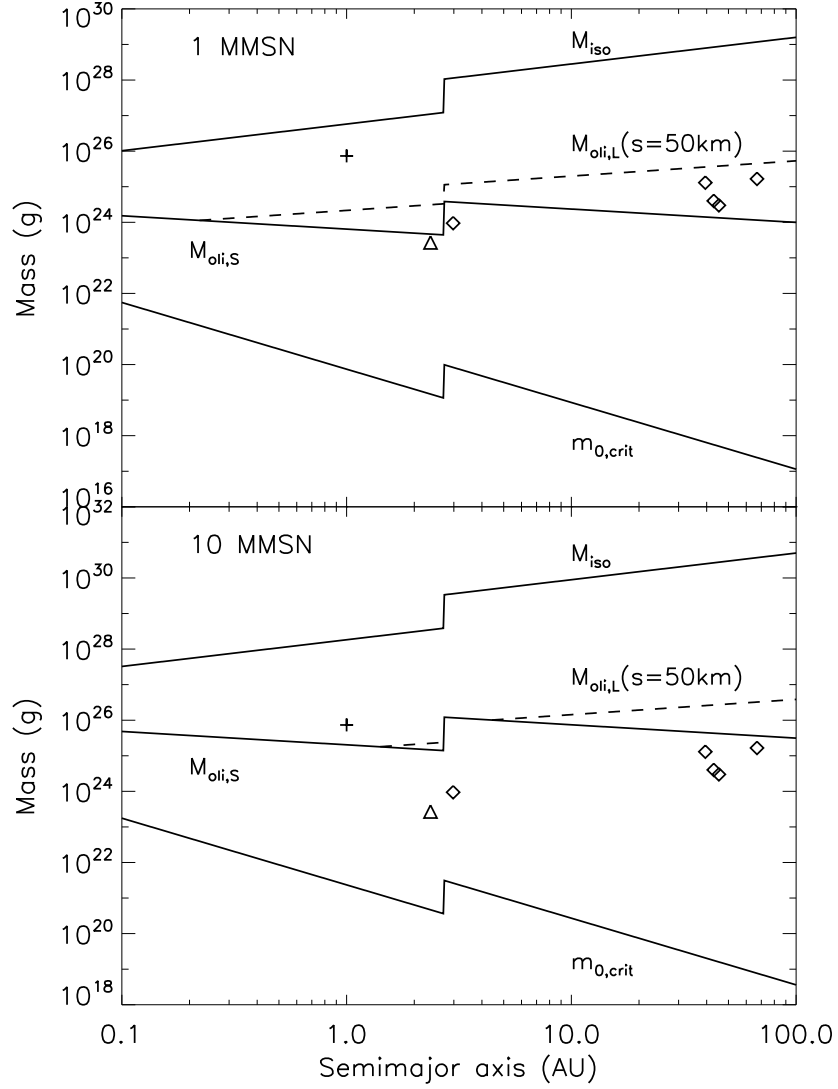}
\end{center}
\caption{Planetary masses at the onset of oligarchic growth $M_{\rm oli,S}$ for $m_0 < m_{\rm 0,crit}$
and $M_{\rm oli,L}$ for $s = $ 50 km and $q = -2.5$. The isolation mass $M_{\rm iso}$
and the critical planetesimal mass $m_{\rm 0, crit}$ are also shown.
The upper and lower panels show the cases of $f_{\sigma} = 1$ and 10, respectively.
The diamonds indicate the dwarf planets (Ceres, Pluto, Haumea, Makemake, and Eris in ascending order of $a$).  
The plus and the triangle indicate Moon and Vesta.}
\label{fig:moli2}
 \end{figure*}

Figure~\ref{fig:moli2} shows the planetary mass at the onset of oligarchic growth  
as a function of semimajor axis for the surface densities of 1 and 10 times of the MMSN. 
The overall growth of planets is described as follows.
Runaway growth first commences after the mass of the largest body $M$ becomes more massive than about 
10 times the initial mass of planetesimals $m_0$.
During runaway growth, $u/v_{H}$ decreases with increasing $M$.
If $m_0 < m_{\rm 0,crit}$ [Eq.~(\ref{eq:m0c})], the trans-Hill stage appears 
when $u/v_{H}$ reaches near unity ($\alpha^{-1/8}$).
The mass at the onset of the trans-Hill stage is given by $M_{\rm tran}$ [Eq.~(\ref{eq:mtrans})].
The trans-Hill stage stage ends and oligarchic growth commences when $M$ reaches to $M_{\rm oli,S}$ [Eq.~(\ref{eq:molis2})], 
which is independent of $m_0$. 
If $m_0 > m_{\rm 0,crit}$,  on the other hand, oligarchic growth commences directly after runaway growth. 
In this case, the planetary mass at the onset of oligarchic growth is given by  $M_{\rm oli,L}$  [Eq.~(\ref{eq:molil})], which is proportional to 
$m_0^{3/7}$ for $q = -2.5$. 
When $M = M_{\rm oli,L}$, $u/v_{H}$ takes its minimum ($> \alpha^{-1/8}$) given by  Eq.~(\ref{eq:uminl}).
In Fig.~\ref{fig:moli2}, the dashed line shows $M_{\rm oli,L}$ for $s = $ 50 km. 

Regardless of $m_0$,  $u/v_{H}$ increases with $M$ during oligarchic growth.
Oligarchic growth ends when planets sweep up all surrounding planetesimals and the planetary 
mass at that time is the isolation mass given by 
$M_{\rm iso}$  [Eq.~(\ref{eq:miso})]. 
Before planetary accretion completes, largest remnant planetesimals which are less massive than $\sim M_{\rm oli}$ have a power-law exponent of 
$q \simeq -2.0$ for $m_0 < m_{\rm 0, crit}$ and $q \simeq -2.7$ to $-2.5$ for $m_0 > m_{\rm 0, crit}$. 

\subsection{Dwarf planets in asteroid and Kuiper belts}
In Fig.~\ref{fig:moli2}, we overplot the masses of Vesta \citep{Russell2012} and the five dwarf planets recognized by the IAU as of January 2016: 
Ceres \citep[and references therein]{Thomas2005}, Pluto \citep{Stern2015},  Haumea \citep{Ragozzine2009},
Makemake \citep[we assumed $\rho = 2.3$ g cm$^{-3}$]{Brown2013}, and Eris \citep{Brown2007}.
Their masses are found to be comparable to $M_{\rm oli}$ 
for the solid surface density of the MMSN in a reasonable range of $s$.
This implies that 
they are likely to be the largest remnant planetesimals that failed to become 
planets while large oligarchic bodies ($m > M_{\rm oli}$) were efficiently removed through collisional 
merging with planets during oligarchic growth.
This argument is probably applied to asteroids.
For KBOs,  their accretion might be simply incomplete due to a long accretion timescale, 
although Pluto and Eris might have entered oligarchic growth 
as indicated from their separations 
from a continuous size distribution \citep{Fraser2014,Adams2014}.

In the following, we will have additional discussions for their accretion histories constrained from the mass distribution slopes.
The power-law exponent $q$ for the mass distribution that we use is given by 
$q = 5\alpha_{\rm H}/3 +1$,  where 
$\alpha_{\rm H}$ is the power-law exponent for the differential 
distribution of absolute luminosity function (H-distribution), provided that
the surface albedo and the internal density are independent of size. In the following, we convert the observational $\alpha_{\rm H}$ to 
$q$ using this relationship.  
We do not discuss the mass distribution of objects smaller than $\sim$ 50 km 
in radius as they are likely to be collisional fragments \citep{Bottke2005,Campo2012}.

\subsubsection{Asteroid belt}
\begin{figure*}
\begin{center}
\includegraphics[width=0.7\textwidth]{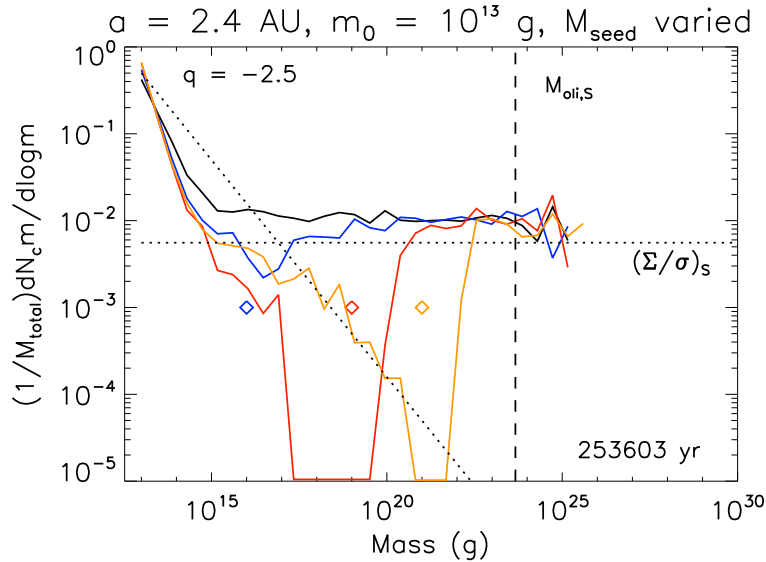}

\end{center}
 \caption{Mass distribution for simulations at $a = 2.4$ AU and $m_0 = 10^{13}$ g with 
asteroid seeds.  The initial mass fraction of asteroid seeds is 0.1 \% (shown as diamonds) and 
the mass of each seed is $10^{16}$ g (blue),   $10^{19}$ g (red), and $10^{21}$ g (orange). 
The black solid curve shows the case without seeds.
The initial inner and outer edges of the planetesimal disk are 2.16 and 2.64 AU. 
We adopt  $f_{\sigma} = 1$, $N_{\rm tr} = 10000$, and $\delta t = 18$ days.}
\label{fig:ast}    
\end{figure*}

Main asteroids have a power-low exponent of $q \simeq -2.0$ for $m  > 10^{21}$ g ($s \simeq$ 50 km) 
while a shallower slope is seen for less massive asteroids \citep{Jedicke2002}. 
The size at the bump is generally considered to be the initial size of planetesimals 
\citep{Morbidelli2009}.  
If planetary accretion starts from equal-mass planetesimals with $s \sim 50$ km, however,
large remnant planetesimals ($m < M_{\rm oli,L}$) turn out to have a mass distribution slope 
of $q \simeq -2.5$, which is steeper than the observed slope. This occurs, since $m_0 > m_{\rm 0,crit}$.  
If small initial planetesimals with $m_0 $ lower than $m_{\rm 0,crit}$ are employed,
the shallow slope ($q \simeq -2.0$) can be reproduced, but not the bump at $10^{21}$ g.
To solve this problem, \citet{Morbidelli2009}
proposed that the size distribution profile of large asteroids up to Ceres was directly inherited from planetesimal formation.

\citet{Weidenschilling2011} performed simulations similar to \citet{Morbidelli2009} while expanding a range of parameters. 
He found that both the bump at $10^{21}$ g  and the shallow slope of large asteroids 
can be reproduced if the initial planetesimal size $s $ is as small as 
100 m. The bump at $10^{21}$ g was not reproduced in our simulation at 1 AU with $m_0 = 10^{13}$ g ($s \simeq 100$ m) even adopting 
$N_{\rm tr} = 10^5$ and similar results are seen in additional simulations at $a = 2.4$ AU (Fig.~\ref{fig:ast}).
This may indicate that collisional fragmentation and gas drag taken into account by \citet{Weidenschilling2011} but 
ignored in our simulations are important or that resolutions of our simulations are still insufficient.

Nevertheless, we find that outcome similar to simulations of  \citet{Weidenschilling2011} can be reproduced 
by introducing a small mass fraction of asteroid seeds while keeping small initial planetesimals.
Figure~\ref{fig:ast} shows the mass distributions at the end of simulations with varying the mass of each asteroid seed $M_{\rm seed}$.
We find that  the mass distribution for $M_{\rm seed} \simeq 10^{19}$ g matches with the observation as a bump (or a shoulder in our plot) 
is produced at $\sim 100 M_{\rm seed}$.
Once the mass fraction of large bodies reaches the theoretical prediction $(\Sigma/\sigma)_{\rm S}$ due to their growth, 
the nature of their subsequent growth is very similar to that seen in the simulation starting with equal-mass planetesimals.
This naturally produces the mass distribution of $q \simeq -2.0$.
The important parameter to reproduce the mass distribution of asteroids in our simulations (Fig.~\ref{fig:ast})
is $M_{\rm seed}$, not $m_0$. Even for lower $m_0$, 
the same bump mass and the same slope for large bodies are expected.
The simulations of \citet{Weidenschilling2011} and ours imply that the fundamental point to reproduce the bump is 
formation of massive runaway bodies before they start to gravitationally stir small planetesimals.

\subsubsection{Kuiper belt}

\begin{figure*}
\begin{center}
\includegraphics[width=0.65\textwidth]{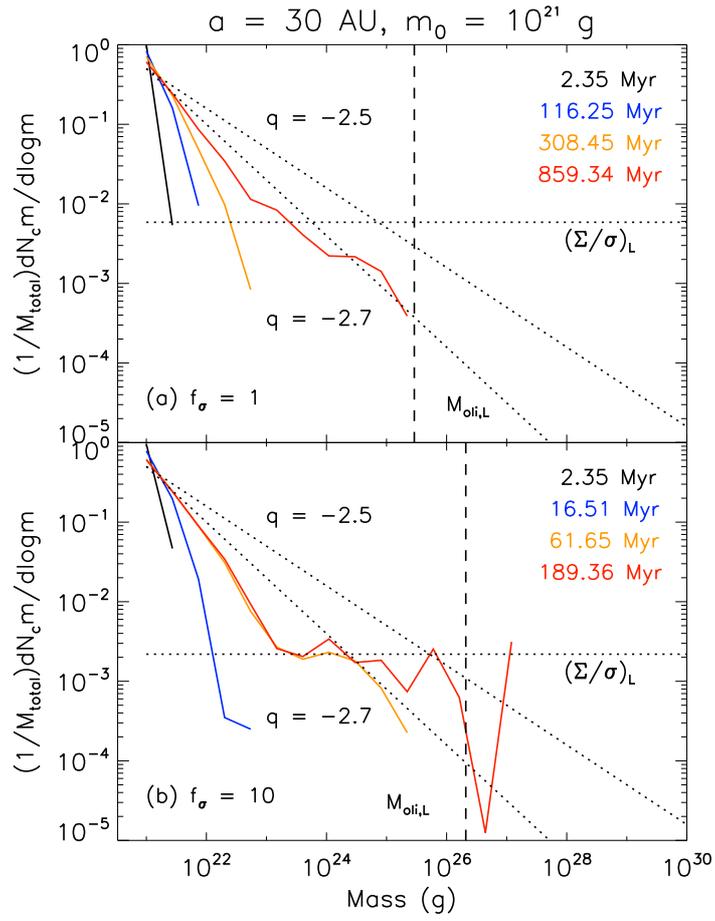}
\end{center}
\caption{Mass distribution for $a = 30$ AU and $m_0 = 10^{21}$ g. The surface density factor $f_{\sigma}$ is (a) 1 and (b) 10. 
The inclined dotted lines are slopes of $q = -2.5$ and -2.7.}
\label{fig:kbo}
\end{figure*}

KBOs are classified into the dynamically excited hot population with large orbital inclinations and eccentricities
and the cold population with low inclinations and eccentricities \citep{Bernstein2004}.
Similarly to asteroids, the both hot and cold populations of KBOs have the bump at $m \sim 10^{21}$ g in their 
mass distributions. 
The power-law exponents for bodies more massive than the bump mass
are $q = -2.45^{+0.32}_{-0.11}$ for the hot population
and $q = -3.5^{+0.3}_{-0.7}$ for the cold population \citep{Fraser2014}. 
The slopes of the absolute brightness magnitudes \citep{Petit2011,Adams2014,Fraser2014}
with corrections due to distances of KBOs are steeper than those of  the apparent magnitudes  
\citep{Bernstein2004,Fraser2010}.
\citet{Fraser2014} found that 
after taking into account the bimodal albedo distribution of hot KBOs, their size distribution 
is very similar to that for  Jupiter's trojans ($q = -2.67 \pm 0.33$). 
This agreement strongly supports the idea of the Nice model that 
trojans were captured during dynamical instability of the giant planets \citep{Morbidelli2005}.
Hot KBOs are likely to have been injected to the current orbits during 
the instability \citep{Malhotra1993,Hahn2005,Levison2008,Lykawka2008,Nesvorny2012}.

The mass distribution slope and the bump mass of hot KBOs can be naturally 
explained if  $m_0$ is the bump mass  as 
$m_0  > m_{\rm 0,crit}$.
Figure~\ref{fig:kbo} shows the mass distribution for the simulation with $m_0 = 10^{21}$ g at $a = 30$ AU.
A simulation with $f_{\sigma} = 1$ is performed in addition to the case with $f_{\sigma} = 10$.
The slope from the simulation is consistent with the observed slope $q \simeq -2.7$ after the albedo correction, 
particularly if $f_{\sigma} = 1$. 
The timescale for the largest body's mass to reach the mass of Pluto  ($1.3 \times 10^{25}$ g) is 
60 Myr for $f_{\sigma} = 10$. If KBOs grew for $\sim$ 700 Myr as suggested by the Nice model \citep{Gomes2005},
a lower surface density down to $f_{\sigma} \sim 1$ is acceptable.  
Our simulations are rather local while KBOs might have formed in a wide range of $a$.
In this case, the mass distribution of KBOs is a superposition of those from different regions. 
Such a scenario was examined by \citet{Kenyon2012}. 
Their results still imply that the steep slope and the bump at $10^{21}$ g are difficult to be reproduced 
if $m_0$ is lower than  $m_{\rm 0, crit}$ (see their Fig.~14). 

Formation of cold KBOs by standard collisional merging is problematic because of their long growth timescale. 
The blue curve at $t = $16.51 Myr in Fig.~\ref{fig:kbo}b is similar to the slope of cold KBOs.
Since the current cold population has $f_{\sigma} \sim 10^{-5}$ \citep{Fraser2014}, the timescale to produce 
the blue curve is $\sim 10^{13}$ yr or 3-4 orders of magnitude longer than the age of the solar system even at 30 AU
(while cold KBOs locate at $a \sim 45$ AU). 
The size distribution of cold KBOs seems primordial reflecting their formation from 
dust. They probably formed in situ and experienced few subsequent collisions 
\citep{Batygin2011,Parker2012}.



\subsection{Planetesimal-driven migration}

\citet{Minton2014} showed five conditions
for planetesimal-driven migration (PDM) to be triggered.
Using our symbols, these conditions are expressed as  
(1) $\sigma/\Sigma > 3$,
(2) $M/M_{\rm stir} > 100$,
(3) $u/v_{\rm H} < 5$,  
(4) $M \ge 5M_{\rm near}$ and
(5) $t_{\rm mig} < (d\log R/dt)^{-1} $.
Here $M_{\rm near}$ is the largest mass of nearby protoplanets in the scattering zone 
of the protoplanet of interest and  
 $t_{\rm mig}$ is the migration timescale.
The condition (4) is probably replaced by $M > M_{\rm oli}$ as massive nearby 
protoplanets inevitably exist ($b_{\rm e} < 3$) before the onset of oligarchic growth.
The condition (5) represents a constraint that 
the protoplanet does not encounter with  
other massive protoplanets during PDM. 
The conditions (1) and (5) are fulfilled at large $a$.
The present study shows that the conditions (2), (3), and (4) are 
simultaneously fulfilled when $M \sim M_{\rm oli}$.
Minton and Levison (2014) examined when all the conditions are fulfilled 
by performing simulations of planetary accretion around 1 AU.
The masses of migration candidates were found to be 0.1 to 10 times the lunar mass 
for $f_{\sigma} = 1.1-4.5$ and $s = 25-50$ km. 
This lowest mass of migration candidates is close to $M_{\rm oli}$ estimated by the present study (Fig.~\ref{fig:moli2}).
Therefore, PDM is likely to occur during early to middle oligarchic growth before
the largest protoplanet is caught up by other nearby protoplanets in mass and at large $a$ ($ > $ 1AU).
Our study also showed that the condition (3) can be better fulfilled for smaller $m_0$, 
consistent with the finding of \citet{Minton2014}.

PDM is not important in our simulations 
at 0.1 and 1 AU (Fig.~\ref{fig:avsem}).
On the other hand, we see embryos migrate back and
force between the inner and outer edges of the 
planetesimal annulus and migration enhances mutual merging between embryos 
for the simulations at $a =$ 30 AU (Fig.~\ref{fig:avsem}d). 
The most massive embryos ($> 10^{27}$ g) 
eventually migrate inward leaving only moderately large embryos in the original location 
of the planetesimal annulus. 
Actual effects of PDM on planetary accretion are under investigation 
adopting wide planetesimal disks (but see Figs.~15 and 16 of \citet{Morishima2015}).

\subsection{Effects of initial mass distribution}

\begin{figure*}
\begin{center}
\includegraphics[width=0.7\textwidth]{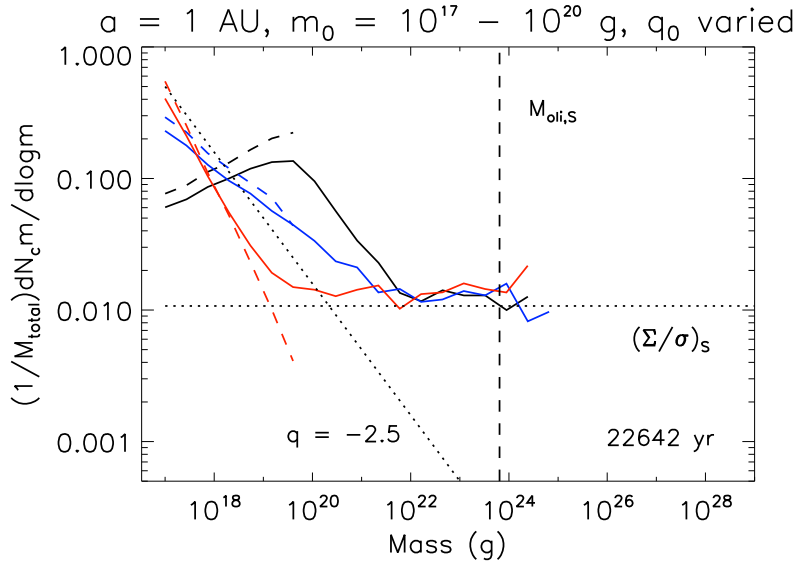}
\end{center}
\caption{Effects of initial mass distribution for $a = 1$ AU and $f_{\sigma} = 1$. 
Initial planetesimal masses range between $10^{17}$ and $10^{20}$ g
and the initial power-law index $q_0$ is $-1.8$ (black), $-2.3$ (blue), and -2.8 (red).
For each case,  the initial and final mass distributions (at  $t = 0$ and 22642 yr) are shown by 
dashed and solid curves. The initial velocity dispersion is the escape velocity of the smallest body.}
\label{fig:imd}
\end{figure*}

We have only considered planetary accretion starting with equal-mass planetesimals, except for the simulations shown in Fig.~\ref{fig:ast}.
It is highly likely, however, that initial planetesimals have an extended mass distribution.
We discuss its effect but limit to the case where
an initial planetesimal mass distribution is given 
by a single power-law with the exponent $q_0$, the upper and lower cut-off masses, $m_{\rm 0U}$ and $m_{\rm 0L}$.

Since $q$ increases up to $\sim -2.5$ during runway growth for the case of initially equal-mass planetesimals,
a similar result is expected as long as $q_0 < -2.5$.
For $q_0 > -2.0$,  on the other hand,
both the mass and stirring power concentrate 
near $m_{\rm 0U}$. 
Thus, accretion proceeds in a similar fashion to a case of initial equal-mass planetesimals
with $m_0 \sim m_{\rm 0U}$ while the initial mass slope for $m < m_{\rm 0U}$ is roughly retained. 
Indeed, these predictions are found to be true in 
simulations (Fig.~\ref{fig:imd}). The mass distribution slope of $q \sim -2.5$ during runway growth
is realized for all bodies for $q_0 < -2.5$ whereas this slope is seen only for bodies with $m > m_{\rm 0U}$ for $q_0 > -2.0$.

It is difficult to predict the evolution of the mass spectrum if $ -2.5 < q_0 < -2.0$.
For a specific case with $q_0 = -2.3$ (Fig.~\ref{fig:imd}), 
planetary accretion proceeds with nearly retaining a single power-law mass distribution during runway growth
even for bodies more massive than $m_{\rm 0U}$.
It is unclear whether a single power-law is always realized for other cases.  
However, we assume it for the discussion below.

The theory described in Section~2 is probably extended to the case with an initial mass distribution, as follows. 
If the trans-Hill stage appears, as is the case for all three simulations shown in Fig.~\ref{fig:imd},
the planetary mass $M_{\rm tran}$ at the onset of the trans-Hill stage is given by 
Eq.~(\ref{eq:mtrans}) but with $q$ and $m_0$ modified as follows:
\begin{eqnarray}
q &=& -2.5,  \hspace{0.3em} m_0 = m_{\rm 0L} \hspace{0.5em} ({\rm for}  \hspace{0.5em} q_0 < -2.5), \nonumber \\
q &=& q_0,  \hspace{0.3em} m_0 = m_{\rm 0L} \hspace{0.5em} ({\rm for}  \hspace{0.5em} -2.5 < -q_0 < -2.0),\\
q &=& -2.5,  \hspace{0.3em} m_0 = m_{\rm 0U} \hspace{0.5em} ({\rm for}  \hspace{0.5em} q_0 > -2.0).  \nonumber
\end{eqnarray}
If $M_{\rm tran}$ for $ -2.5 < q_0 < -2.0$ is larger than $M_{\rm tran}$ for $q_0 > -2.0$,
the former is probably replaced by the latter.
As long as the trans-Hill stage appears, the planetary mass 
at the onset of oligarchic growth is given by $M_{\rm oli,S}$, as repeatedly discussed.
If the initial planetesimals are massive and oligarchic growth directly commences after runway growth,
$M_{\rm oli,L}$ is given by  Eq.~(\ref{eq:molilq}) but again $q$ and $m_0$ are modified as shown above.
To repeat, the modification for $-2.5 < q_0 < -2.0$ is speculative and more simulations or advanced theories are necessary.

\subsection{Effects of damping forces}
Collisional damping (or fragmentation) and gas drag are known to be important during oligarchic growth 
\citep{Kokubo2000,Chambers2008,Kobayashi2010}.
We discuss whether these effects are important already at the onset of oligarchic growth and, if so, 
whether the planetary mass $M_{\rm oli}$ is modified by these effects.

\subsubsection{Collisional damping} 
The velocity change rate due to collisional damping is give as \citep{Goldreich2004,Lithwick2014}
\begin{equation}
\left(\frac{1}{u}\frac{\mathrm{d}u}{\mathrm{d}t}\right)_{\rm coll} = \frac{\sigma \Omega}{\rho s}. 
\end{equation}
The equilibrium velocity determined by the balance between viscous stirring of large bodies [Eq.~(\ref{eq:dudt})]
and collisional damping is given as 
\begin{equation}
\left(\frac{u}{v_{\rm H}}\right)_{\rm coll} = \left(\frac{s \Sigma}{R \sigma} \alpha^{-2} \right)^{p_c} =  \left(\frac{\rho s R_{\rm H}}{a \sigma}\right)^{p_c},
\end{equation}
where the power-law index $p_c$ is 1/4 for $u > v_{\rm H}$ and 1 for $u \le v_{\rm H}$
and we used the equation for oligarchic growth
$\Sigma a = RR_{\rm H}\rho \alpha^2$ [Eq.~(\ref{eq:mem})] with dropping off order of unity factors in the second equality.
If $(u/v_{\rm H})_{\rm coll}$ is lower than $(u/v_{\rm H})_{\rm oli}$ [Eq.~(\ref{eq:uoli})],
collisional damping cannot be ignored at the onset of oligarchic growth or in the earlier stages.
In Fig.~\ref{fig:damp}, we plot $(u/v_{\rm H})_{\rm coll}$ derived using $R_{\rm H}$ for the oligarchic body with a mass of $M_{\rm oli}$
and compare it with $(u/v_{\rm H})_{\rm oli}$. Collisional damping is found to be important 
for small $s$ and $a$, or $s < R \alpha^{3/4}$. 

Two possibilities were indicated for evolution of $\Sigma/\sigma$ during the trans-Hill stage in the collisional regime ($s < R \alpha^{3/4}$).
If $(u/v_{\rm H})_{\rm coll}$ is retained to be about unity \citep{Lithwick2014},  $\Sigma/\sigma = \alpha^{2}R/s$.
On the other hand, if equal accretion is realized in the regime of $\alpha^{1/2} < u/v_{\rm H} < 1$ \citep{Shannon2016}, 
$(u/v_{\rm H})_{\rm coll} = \alpha^{-3/4}(s/R)^{1/2}$ and $\Sigma/\sigma = \alpha^{4/5}(R/s)^{1/2}$.
In either case, $\Sigma/\sigma$ for the collisional regime is expected to increase with $R$ and 
eventually exceed that for the collisionless regime $(\Sigma/\sigma)_{\rm S}$ [Eq.~(\ref{eq:sig0s2})].
Thus, $M_{\rm oli}$ for the collisional regime is expected to be larger as well.
 
Collisional damping is automatically included in all of our simulations (Section~3), 
and it should be particularly important for the simulation 
with $a = 0.1$ AU and $m_0 = 10^{13}$ g. In this simulation, 
we see that $\Sigma$ indeed increases with $R$ during the trans-Hill stage but 
much more gradually than the theoretical prediction (see Fig.~\ref{fig:mdf}a for $m_0 = 10^{17}$ g 
as this case gives a slope similar to that for $m_0 = 10^{13}$ g). 
The mass $M_{\rm oli}$ also agrees with that for the collisionless regime rather than that for the collisional regime.
The discrepancy is caused by a fact that significant depletion of the smallest bodies occurs in simulations 
while the theory assumes a fixed $\sigma$.
As the surface density of the smallest bodies decreases, so is the 
the effect of collisional damping. 

The effect of depletion of the smallest bodies is expected to be relatively unimportant at large $a$ or small $\alpha$ although 
collisional damping at large $a$ is only important for very small $s$.
\citet{Shannon2016} performed simulations for formation of cold KBOs from cm-sized particles 
and showed that equal accretion is actually realized in the collisional case as their theory predicted.


\subsubsection{Gas drag}
The effect of gas drag is expected to play a similar role to collisional damping.
The velocity change due to gas drag in the quadratic regime is given as 
\begin{equation}
\left(\frac{1}{u}\frac{\mathrm{d}u}{\mathrm{d}t}\right)_{\rm gas} = \frac{\rho_{\rm gas} u}{\rho s}, 
\end{equation}
where $\rho_{\rm gas}$ is the gas density.
The equilibrium velocity determined by the balance between viscous stirring of large bodies and gas drag is 
\begin{equation}
\left(\frac{u}{v_{\rm H}}\right)_{\rm gas} = \left(\frac{s \Sigma}{R R_{\rm H} \rho_{\rm gas}} \alpha^{-2} \right)^{p_g} =  \left(\frac{\rho s}{a \rho_{\rm gas}}\right)^{p_g},
\end{equation}
where the power-law index $p_g$ is 1/5 for $u > v_{\rm H}$ and 1/2 for $u \le v_{\rm H}$ and we again used the equation for oligarchic growth
with dropping off order of unity factors in the second equality. 
In Fig.~\ref{fig:damp}, we plot $(u/v_{\rm H})_{\rm gas}$ assuming the gas density for the MMSN
$\rho_{\rm gas} = 1.4 \times 10^{-9} (a/{\rm 1\hspace{0.3em}AU})^{-11/4}$ g cm$^{-3}$.
Gas drag is found to be important for small $s$ as expected and also for large $s$.

For the case of small $m_0$, 
we expect $\Sigma/\sigma \propto R^{2/3}$ during the trans-Hill stage, 
if equal accretion is realized in the regime of $\alpha^{1/2} < u/v_{\rm H} < 1$. 
Here, we ignored mass depletion due to radial migration of small bodies.
As well as the case with collisional damping, the expected value of $\Sigma/\sigma$ and $M_{\rm oli}$ are 
larger than those for the non-damping case.

 

The condition for $(u/v_{\rm H})_{\rm gas} < (u/v_{\rm H})_{\rm oli}$ is also written 
as  St $< \alpha^{-1/(8p_g)}/e$ ($\sim 1000$ at the onset of oligarchic growth), 
where St $\simeq \rho s/(\rho_{\rm gas} a e)$ is 
the Stokes number for the quadratic gas drag regime. 
For St $< 100$,  the accretion rate of a large body we adopted  [Eq.~(\ref{eq:dr1dt})] is no longer
valid since small bodies'  orbits in the Hill sphere of a large body are highly modified by gas drag.
The planetary accretion in this regime, called pebble accretion, is generally very efficient 
\citep{Ormel2010c,Lambrechts2012,Morbidelli2012}.
This effect was successfully studied for formation of giant planets  
\citep{Lambrechts2014,Morbidelli2015,Levison2015a} and
terrestrial planets \citep{Levison2015b,Sato2016}.
This mechanism is also able to reproduce the mass distribution of asteroids
and potentially that for KBOs \citep{Johansen2015b,Johansen2015a}.  

For the case of large $m_0 ( > m_{\rm 0,crit})$, 
to derive $M_{\rm oli, L}$, we assumed a single power-law for the mass distribution.
The power-law index $q$ always takes  $\simeq -2.5$ whether gas drag is included or not
\citep{Kokubo2000,Morishima2013}.
Thus, $M_{\rm oli, L}$ is probably not modified by gas drag, although it can accelerate planetary accretion.

\section{Summary}

Oligarchic growth commences when the velocity dispersion $u$ relative to 
the Hill velocity $v_{\rm H}$ of the largest body takes its minimum.
We found that if the initial planetesimal mass $m_0$ is small enough, 
$u/v_{\rm H}$ becomes as low as unity during 
the intermediate (trans-Hill) stage between the runaway and oligarchic growth stages. 
In this case, $M_{\rm oli}$ is independent of $m_0$.
If  $m_0$ is large, on the other hand, oligarchic growth commences directly after 
runaway growth, and $M_{\rm oli} \propto m_0^{3/7}$.
We found that the contribution of mutual merging between large bodies to their growth is 
comparable to or slightly larger than that due to merging with small planetesimals at any stage of planetary accretion.

The planetary mass $M_{\rm oli}$ for the solid surface density of the Minimum Mass Solar Nebula is close to the masses of the dwarf planets
in a reasonable range of $m_0$. This indicates that they are likely to be the largest remnant planetesimals that failed to become planets.
Bodies more massive than $M_{\rm oli}$ probably merged with massive protoplanets once existed.
For KBOs, alternatively, their accretion might have been externally halted by orbital instability of giant planets 
when the largest KBOs reached $M_{\rm oli}$ in mass.

The mass distribution bump at $10^{21}$ g ($s \sim $ 50 km) and the slope,  $q \simeq -2.7$, of hot KBOs
are reproduced if $m_0$ is the bump mass.  
On the other hand, the small initial planetesimals, $m_0 \sim 10^{13}$ g or less, are favored for
reproducing the slope of asteroids, $q \simeq -2.0$, while the bump at 
$10^{21}$ g can be reproduced by introducing a small number of 
asteroid seeds each with mass of $10^{19} $ g (10 km in radius).
The present study rather focused on the basic theory that connects between the initial and the final mass distributions.
Detailed comparison between modeled and observed mass distributions should be made somewhere else.

\section*{Acknowledgements}
We thank the reviewers, Yoram Lithwick and Eiichiro Kokubo, for their constructive comments that
helped us to improve the manuscript.
This research was carried out in part at the Jet Propulsion Laboratory, California Institute of Technology, 
under contract with NASA. Government sponsorship acknowledged. 
Simulations were performed using a JPL supercomputer, Aurora.
 
\bibliographystyle{model2-names}
\bibliography{trans} 

\end{document}